\pdfoutput=1
\documentclass[11pt,onecolumn]{cleantechnicalreport}

\usepackage[authoryear,sort&compress,round]{natbib}
\usepackage{hybridfrontmatter}
\usepackage{wrapfig}
\usepackage{placeins}
\usepackage{tikz}

\title{Can KV Cache Be Reused Across Model Updates? \\
A DeepSeek-V4-Flash Case Study
}

\newcommand{\shorttitle}{Can KV Cache Be Reused Across Model Updates? A DeepSeek-V4-Flash Case Study}
\newcommand{\SciTixEmailFont}{\normalfont\fontsize{9}{11}\selectfont}
\newcommand{\SciTixBrandFont}{\normalfont\bfseries\fontsize{16}{19}\selectfont}
\newcommand{\SciTixAuthorCell}[2]{%
  \begin{tabular}[t]{@{}c@{}}
    \ReportAuthorName{#1}\\[0.2mm]
    {\SciTixEmailFont #2}%
  \end{tabular}%
}

\makeatletter
\gdef\@author{%
  \begin{tabular}{@{}c@{\hspace{9mm}}c@{\hspace{9mm}}c@{}}
    \SciTixAuthorCell{Zhikun Tao}{} &
    \SciTixAuthorCell{Fangyong Jin}{} &
    \SciTixAuthorCell{Junping Zhao\footnotemark[1]}{} \\[8mm]
    \multicolumn{3}{c}{%
          \SciTixBrandFont\color{ReportTheme}Rank Computing{\color{ReportAccent}}%
    }%
  \end{tabular}%
}
\makeatother


\begin{abstract}
Reusing key--value (KV) caches across language model updates could avoid repeated computation of shared context, but changes in model weights raise a fundamental question about what these caches preserve. We investigate direct cache reuse across four model updates, with a detailed study of DeepSeek-V4-Flash to Flash-0731. Our central finding is that caches can retain useful task information without faithfully reproducing the updated model's behavior. Retrieval accuracy varies substantially across updates, ranging from near-native performance to large losses. 
With source caches already available, reuse achieves median time-to-first-token speedups of $2.10$--$5.08\times$ on three longer-prefix QA benchmarks.
To characterize the behavioral differences behind these outcomes, we introduce Functional Cache Divergence, which compares next-token distributions under reused and native caches. 
Direct KV cache reuse better preserves native model behavior in natural-text continuation, whereas compatibility degrades with increasing prefix length in single-token retrieval tasks. 
Answer-margin analyses help explain how correct retrieval decisions survive substantial internal state changes. Yet even agreement on the first generated token does not ensure agreement in answer content or stopping behavior. These findings establish task utility and fidelity to native inference as distinct criteria for evaluating cross-version KV cache reuse.
\end{abstract}

\begin{document}

\begingroup
  \renewcommand{\thefootnote}{\fnsymbol{footnote}}
  \maketitle
  \footnotetext[1]{Corresponding author: \href{mailto:jason@siflow.cn}{jason@siflow.cn}}
\endgroup

\noindent\begin{minipage}{\linewidth}
  \setlength{\parskip}{0pt}
  \centering
  \includegraphics[width=\linewidth]{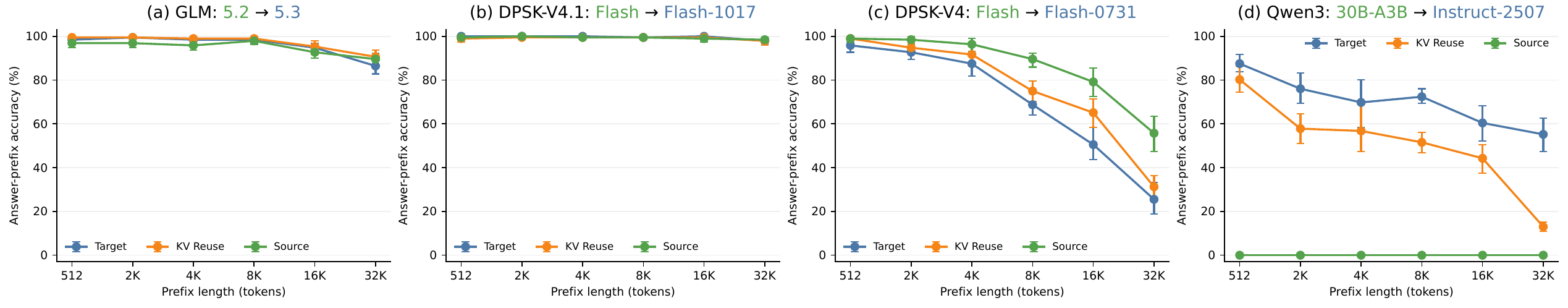}
  \captionsetup{hypcap=false}
  \captionof{figure}{
  Needle-in-a-Haystack (NIAH)~\citep{hsieh2024ruler} retrieval accuracy across four model pairs and prefix lengths in the original sweep. Each prefix contains lots of key-value records, with each key and value represented by a single token. The model is asked to return the value for a given key, testing its ability to retrieve specific information from long contexts. Panel titles indicate source → target. Source and Target use their own prefix KV caches, while KV Reuse denotes target-model inference using the source model's prefix KV cache.
  }
  \label{fig:cross-version-overview}
\end{minipage}
\par
\clearpage

\section{Introduction}
Large language models (LLMs) are increasingly refined through successive rounds of post-training, allowing improved versions to be released rapidly without changing their core architecture or size.  
For example, DeepSeek-V4-Flash-0731 retained the architecture and size of its preview model and improved agent capabilities through further post-training, roughly three months after the preview's release on April 24, 2026~\citep{deepseekai2026flash0731}. GLM-5.3 similarly builds on GLM-5.2's base model, with improvements in coding and agent tasks attributed to post-training~\citep{zai2026glm53}.

Even as models evolve, applications often continue to rely on the same context. This includes shared system prompts, frequently requested prompts, retrieved documents reused across queries in retrieval-augmented generation (RAG), and common instructions or examples for recurring data-processing tasks. 
In these long conversations and repeated document queries, key--value (KV) caches avoid recomputing previously processed context. 
A cache created before an update still encodes the context using the old model weights, reusing it could save computation but may change how the updated model uses that context and what it generates.

Prior work explores KV-based communication among agents~\citep{shi2026kvcomm,yun2026prefill}, mapped cache reuse between models in the same family~\citep{heo2026cross,qu2026cachebridge}, and reuse across changing contexts and model checkpoints~\citep{shi2026kvsharearena}.
We focus on \emph{direct KV-cache reuse} across successive model versions with unchanged core architecture and size. The updated model consumes its predecessor's cache without mapping or selective recomputation. We ask: \emph{what remains compatible when a newer model consumes caches produced by an earlier version?}
We evaluate retrieval, answer quality, time to first token (TTFT), and generation behavior using three configurations. \textbf{Target} and \textbf{Source} each use one checkpoint for both prefill and generation. \textbf{KV Reuse} combines source prefill with target continuation.

The four model updates in Figure~\ref{fig:cross-version-overview} show more reliable KV reuse in models with stronger retrieval ability. On single-token needle-in-a-haystack (NIAH) tasks with 512--32K prefixes, GLM-5.2 $\rightarrow$ GLM-5.3 reuse closely tracks Target through 16K and improves 32K accuracy from 86.46\% to 90.62\%. DeepSeek-V4.1-Flash $\rightarrow$ Flash-1017 maintains near-native accuracy, with both Target and KV Reuse reaching 97.92\% at 32K. DeepSeek-V4-Flash $\rightarrow$ Flash-0731 reuse improves Target accuracy at every measured length, although retrieval becomes less accurate as prefixes grow. Qwen3-30B-A3B $\rightarrow$ Instruct-2507 loses accuracy under reuse at every length, with the largest gap at 32K despite RoPE phase alignment.

Our detailed evaluation centers on DeepSeek-V4-Flash $\rightarrow$ Flash-0731. We first compare direct reuse with a ridge mapper adapted to DeepSeek's composite cache, then measure answer quality and TTFT on four long-context QA benchmarks with the source cache prepared. We also inject Flash's caches into both Flash-Base and Flash-0731 to examine transfer toward earlier and later checkpoints. Complete QA responses reveal changes in answer content and stopping behavior.

We connect these outcomes to changes in predictions and internal computation. \emph{Functional Cache Divergence} (FCD) measures the difference between the target model's next-token probability distributions using its own cache and the source cache, with subsequent input tokens and prediction positions held fixed. This scaled Jensen--Shannon divergence captures probability shifts even when the selected token stays the same, revealing distinct patterns for retrieval and natural-text continuation.

Internal analyses explain how retrieval survives state changes. Flash-0731 retains many correct answers through sufficient score margins, while Flash-Base recovers retrieval with higher answer-token scores despite changed states and cache selections. At the three inspected layers, replacing hidden states changed by earlier layers affects attention outputs more than replacing local caches.

This technical report presents four main findings:

\begin{enumerate}
  \item \textbf{Stronger retrieval models preserve correct answers more reliably under KV reuse.}
  GLM-5.3 and DeepSeek-V4.1-Flash-1017 retain high retrieval accuracy when using their predecessors' KV caches. At 32K, their harmful flip rates---the fractions of Target-correct answers that become incorrect under reuse---are 4.22\% and 0.53\%, compared with 26.53\% for V4-Flash and 78.30\% for Qwen. These larger models with stronger retrieval preserve correct answers more reliably. V4.1-Flash reuse also reaches 61.46\% accuracy at a 512K-token prefix (Section~\ref{sec:length-compatibility}).

  \Needspace{6\baselineskip}
  \item \textbf{KV reuse substantially accelerates QA on longer contexts.}
  With the source cache prepared, median TTFT speedups are $5.08\times$ on HotpotQA, $2.10\times$ on 2WikiMQA, and $2.51\times$ on MultiFieldQA-en. MultiFieldQA-en maintains F1, while HotpotQA and 2WikiMQA lose about 3 and 7 F1 points. On BABILong's shorter prefixes, cache-access costs outweigh the saved computation, yielding a $0.86\times$ speedup (Section~\ref{sec:quality-latency}).

  \item \textbf{KV caches carry the source model's processing of context and response instructions, changing how receivers answer.}
  Flash-produced caches improve retrieval in both Base and Flash-0731. With the same instruction prompt, Base typically restates the question when using its own cache and scores 0\% accuracy. Injecting Flash's cache raises its accuracy to 91.15\% at 2K tokens while its weights remain fixed. In natural QA, reuse also changes answer content and stopping, including cases where both runs begin with the same answer token (Sections~\ref{sec:cache-answer-behavior} and~\ref{sec:sequence-agreement}).

  \item \textbf{Changes in predictive distributions grow with context length for retrieval and stay stable for natural-text continuation.}
  In the Flash update, mean NIAH FCD rises from 2.87 at 512 tokens to 30.62 at 32K, while FineWeb-Edu stays within 3.04--3.64 across the same lengths. Replacing the prefix cache increasingly changes retrieval-token probabilities as context grows, while natural-text continuation retains a much lower, flatter divergence profile (Section~\ref{sec:fcd-tasks}).

\end{enumerate}

\section{Experimental Setup and Measurements}
\label{sec:study-design}

We study how a target model continues from a prefix cache produced by another checkpoint in the same model family. The evaluation measures retrieval, answer quality, and latency, then examines complete responses, predictive distributions, and internal computation. DeepSeek-V4-Flash $\rightarrow$ Flash-0731 is the main pair for QA and controlled interventions. Flash-to-Base adds a comparison across training stages, while GLM, V4.1-Flash, and Qwen provide retrieval comparisons across updates.

\subsection{Inference Configurations}
\label{sec:handoff}

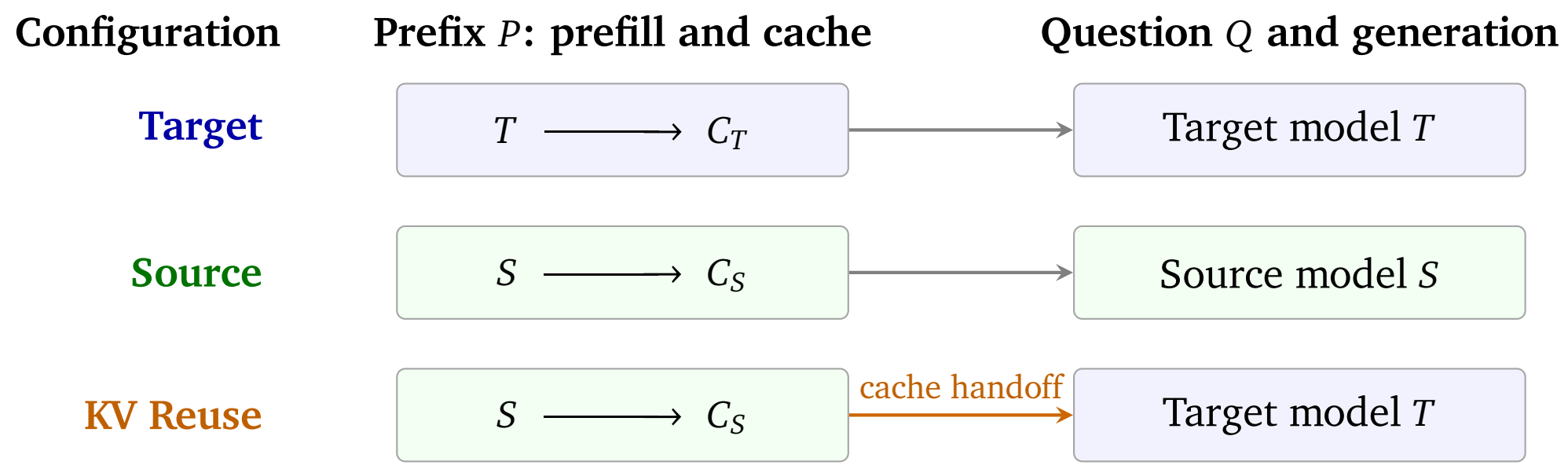
\begin{figure}[ht]
  \centering
  \begin{tikzpicture}[x=1cm,y=1cm,
    block/.style={draw=black!35,rounded corners=2pt,minimum width=3.8cm,minimum height=0.78cm,align=center,font=\small},
    flow/.style={->,>=stealth,thick,draw=black!50}]
    \node[font=\small\bfseries] at (0,0.6) {Configuration};
    \node[font=\small\bfseries] at (4.0,0.6) {Prefix $P$: prefill and cache};
    \node[font=\small\bfseries] at (9.7,0.6) {Question $Q$ and generation};
    \node[anchor=east,font=\small\bfseries,text=blue!65!black] at (1.1,-0.2) {Target};
    \node[block,fill=blue!5] (tc) at (4.0,-0.2) {$T \;\xrightarrow{\hspace{1cm}}\; C_T$};
    \node[block,fill=blue!5] (td) at (9.7,-0.2) {Target model $T$};
    \draw[flow] (tc) -- (td);
    \node[anchor=east,font=\small\bfseries,text=green!45!black] at (1.1,-1.4) {Source};
    \node[block,fill=green!5] (sc) at (4.0,-1.4) {$S \;\xrightarrow{\hspace{1cm}}\; C_S$};
    \node[block,fill=green!5] (sd) at (9.7,-1.4) {Source model $S$};
    \draw[flow] (sc) -- (sd);
    \node[anchor=east,font=\small\bfseries,text=orange!75!black] at (1.1,-2.6) {KV Reuse};
    \node[block,fill=green!5] (rc) at (4.0,-2.6) {$S \;\xrightarrow{\hspace{1cm}}\; C_S$};
    \node[block,fill=blue!5] (rd) at (9.7,-2.6) {Target model $T$};
    \draw[flow,draw=orange!80!black] (rc) -- node[above,font=\scriptsize,text=orange!75!black] {cache handoff} (rd);
  \end{tikzpicture}
  \caption{Three inference configurations on the same prefix and question. Target and Source use one model throughout. KV Reuse combines the source-produced prefix cache with target-model continuation.}
  \label{fig:handoff-configurations}
\end{figure}

Let $S$ denote the source model and $T$ the target model. Each input is divided into a shared prefix $P$ and a question suffix $Q$. Prefilling the same prefix with the two models produces $C_S=\operatorname{Prefill}_S(P)$ and $C_T=\operatorname{Prefill}_T(P)$.
Here, $C$ denotes the complete prefix state required for continuation, including the attention and compression states used by DeepSeek-V4. We compare three configurations (Figure~\ref{fig:handoff-configurations}): \textbf{Target} uses $T$ for both prefill and continuation; \textbf{Source} uses $S$ for both stages; and \textbf{KV Reuse} gives $C_S$ to $T$, which processes $Q$ and generates the response. At output position $j$, their predictive distributions are
\begin{equation}
  \begin{aligned}
    p^{\mathrm{Target}}_j &= p_T(\,\cdot\mid C_T,Q,y_{<j}),\\
    p^{\mathrm{Source}}_j &= p_S(\,\cdot\mid C_S,Q,y_{<j}),\\
    p^{\mathrm{Reuse}}_j &= p_T(\,\cdot\mid C_S,Q,y_{<j}).
  \end{aligned}
  \label{eq:three-configurations}
\end{equation}
For free generation, $y_{<j}$ is the output history of the corresponding configuration. For paired distribution measurements, all configurations being compared receive the same prescribed history. Prompt token IDs and decoding settings are identical within each model pair, and free generation uses greedy decoding.
Table~\ref{tab:cross-version-retrieval} additionally compares \textbf{Ridge Mapper} for Flash $\rightarrow$ Flash-0731: the target continues from $M(C_S)$, where $M$ adapts the closed-form ridge mapping of \citet{heo2026cross} to DeepSeek's composite cache. Appendix~\ref{app:niah} describes calibration and cache adaptation.
Figure~\ref{fig:niah-harmful-flips}(b) applies the same centered ridge approach to V4.1-Flash $\rightarrow$ Flash-1017, with adaptation to its cache representation.


\subsection{Tasks and Datasets}
\label{sec:tasks-and-datasets}

\textbf{For retrieval compatibility,} we use controlled key--value NIAH tasks~\citep{hsieh2024ruler,zhu2026periodic} with prefixes of 512, 2K, 4K, 8K, 16K, and 32K tokens. Each length has eight independent histories and 24 queries per history, distributed near 10\%, 50\%, and 90\% of the prefix. Contexts contain approximately one record per 16 tokens, with single-token keys and values. This controlled retrieval task tests a model's ability to locate information in context and directly measures how KV reuse affects the target's retrieval performance. We also extend the V4.1-Flash update to 64K, 128K, 256K, and 512K, retaining eight histories and 24 queries per history at each added length. Appendix~\ref{app:niah} gives the data construction and generation limits.

\textbf{For task quality and latency,} we evaluate the KV Reuse from DeepSeek-V4-Flash to DeepSeek-V4-Flash-0731 on 585 examples: 300 from BABILong, 100 from HotpotQA, 100 from 2WikiMQA, and 85 from MultiFieldQA-en~\citep{kuratov2024babilong,bai2024longbench,yang2018hotpotqa,ho2020constructing}. We measure answer scores and Time to First Token (TTFT) with the source prefix cache already prepared. Appendix~\ref{app:datasets} details dataset selection, input lengths, and evaluation settings.

\textbf{For behavioral and mechanistic analysis,} we compare selected 4K NIAH state traces for Flash $\rightarrow$ Flash-0731 and Flash $\rightarrow$ Flash-Base, alongside token-score margins on the same 192 sweep queries for each receiver. Controlled interventions use a separate Flash-update cohort of 96 queries from 32 independent 4K contexts, with a further 48 queries from 16 contexts for matched perturbations. Natural-QA diagnostics use 39 HotpotQA and 50 2WikiMQA examples to examine answer changes, generation agreement, and stopping. Each task contributes 20 preselected random examples, and frequency summaries use only these random subsets. Appendix~\ref{app:diagnostic-data} describes the diagnostic cohorts and their selection.

\textbf{For functional cache divergence,} we separately measure the Flash update on the same 1,152 NIAH queries used in its six-length sweep, comparing full predictive distributions at the first answer position. We also evaluate 64 long documents from FineWeb-Edu~\citep{penedo2024fineweb}. The same documents are used at all six lengths, with 32 matched continuation positions per document and length, yielding 12,288 paired positions. Both target runs consume identical subsequent document tokens. Appendix~\ref{app:fcd-protocol} specifies sampling and the matched prediction positions.

Paired runs share input tokens and execution settings within each experiment. Appendix~\ref{app:runtime} gives runtime and statistical settings.

\section{Effectiveness and Behavioral Effects of KV Reuse}
\label{sec:reuse-outcomes}

We first examine retrieval across model updates and prefix lengths, then evaluate the quality and latency of the DeepSeek-V4-Flash handoff on natural-language tasks. We next study how transferred caches affect a receiver's answers, including changes in answer content and stopping.

\subsection{Retrieval Evaluation}
\label{sec:length-compatibility}

\begin{table}[!ht]
  \centering
  \caption{NIAH answer-prefix accuracy (\%; higher is better) across model updates. Each prefix length contains 192 queries from eight histories. Sizes in parentheses denote nominal backbone parameters. Target and Source use their own caches; KV Reuse uses the target model with the source cache; Ridge Mapper maps that cache before target-model continuation.}
  \label{tab:cross-version-retrieval}
  \small
  \setlength{\tabcolsep}{8pt}
  \renewcommand{\arraystretch}{1.12}
  \begin{tabularx}{\linewidth}{l*{6}{>{\raggedleft\arraybackslash}X}}
    \toprule
    & \multicolumn{6}{c}{Prefix length (tokens)} \\
    \cmidrule(l){2-7}
    Configuration & 512 & 2K & 4K & 8K & 16K & 32K \\
    \midrule
    \multicolumn{7}{l}{\textbf{GLM-5.2 $\rightarrow$ GLM-5.3 (750B)}} \\
    Target & 98.44 & 99.48 & 98.44 & 98.44 & 94.79 & 86.46 \\
    Source & 96.88 & 96.88 & 95.83 & 97.92 & 92.71 & 89.58 \\
    \rowcolor{ReportTheme!5}
    KV Reuse & 99.48 & 99.48 & 98.96 & 98.96 & 95.31 & 90.62 \\
    \addlinespace[3pt]
    \multicolumn{7}{l}{\textbf{DeepSeek-V4.1-Flash $\rightarrow$ DeepSeek-V4.1-Flash-1017 (552B)}} \\
    Target & 100.00 & 100.00 & 100.00 & 99.48 & 100.00 & 97.92 \\
    Source & 99.48 & 100.00 & 99.48 & 99.48 & 98.96 & 98.44 \\
    \rowcolor{ReportTheme!5}
    KV Reuse & 98.96 & 99.48 & 99.48 & 99.48 & 99.48 & 97.92 \\
    \addlinespace[3pt]
    \multicolumn{7}{l}{\textbf{DeepSeek-V4-Flash $\rightarrow$ DeepSeek-V4-Flash-0731 (284B)}} \\
    Target & 95.31 & 92.19 & 90.10 & 68.75 & 50.52 & 25.52 \\
    Source & 98.96 & 98.96 & 97.92 & 89.58 & 79.17 & 55.73 \\
    \rowcolor{ReportTheme!5}
    KV Reuse & 98.96 & 96.35 & 93.23 & 75.00 & 65.10 & 31.25 \\
    \rowcolor{ReportTheme!12}
    Ridge Mapper~\citep{heo2026cross} & 98.44 & 92.19 & 92.71 & 76.04 & 59.38 & 40.10 \\
    \addlinespace[3pt]
    \multicolumn{7}{l}{\textbf{Qwen3-30B-A3B $\rightarrow$ Qwen3-30B-A3B-Instruct-2507 (30.5B)}} \\
    Target & 87.50 & 76.04 & 69.79 & 72.40 & 60.42 & 55.21 \\
    Source & 0.00 & 0.00 & 0.00 & 0.00 & 0.00 & 0.00 \\
    \rowcolor{ReportTheme!5}
    KV Reuse & 80.21 & 57.81 & 56.77 & 51.56 & 44.27 & 13.02 \\
    \bottomrule
  \end{tabularx}
  \par\smallskip
  \begin{minipage}{0.95\linewidth}
  \end{minipage}
\end{table}

Table~\ref{tab:cross-version-retrieval} and Figure~\ref{fig:cross-version-overview} compare Target, Source, and KV Reuse on NIAH retrieval from 512 to 32K tokens, with model pairs ordered by backbone size~\citep{glm5team2026glm5,deepseekai2026deepseekv4,deepseekai2026deepseekv41flash}. We use answer-prefix accuracy: a response is correct if its whitespace-trimmed text begins with the requested value, and failed requests score zero. GLM reuse closely tracks Target through 16K and improves accuracy from 86.46\% to 90.62\% at 32K. V4.1-Flash reuse stays within 1.04 percentage points of Target at every length, with both reaching 97.92\% at 32K. V4-Flash reuse improves Target accuracy throughout the sweep, including from 25.52\% to 31.25\% at 32K, although it remains below Source's 55.73\% at that length. Qwen shows the opposite pattern: reuse reduces accuracy at every length, with the largest gap at 32K, where accuracy falls from 55.21\% to 13.02\% even with RoPE phase alignment.

Using the closed-form ridge mapper of \citet{heo2026cross}, adapted to DeepSeek's composite cache, improves retrieval at some lengths but does not consistently outperform direct reuse. At 32K, Ridge Mapper reaches 40.10\% accuracy versus 31.25\% for KV Reuse, a gain of 8.85 percentage points (95\% history-cluster bootstrap interval: $[5.21,12.50]$). At 2K, accuracy instead falls from 96.35\% to 92.19\%, and at 16K from 65.10\% to 59.38\%. Mapping therefore provides a useful long-prefix comparison without establishing a uniform advantage across the sweep.

\begin{figure}[!htbp]
  \centering
  \includegraphics[width=\linewidth]{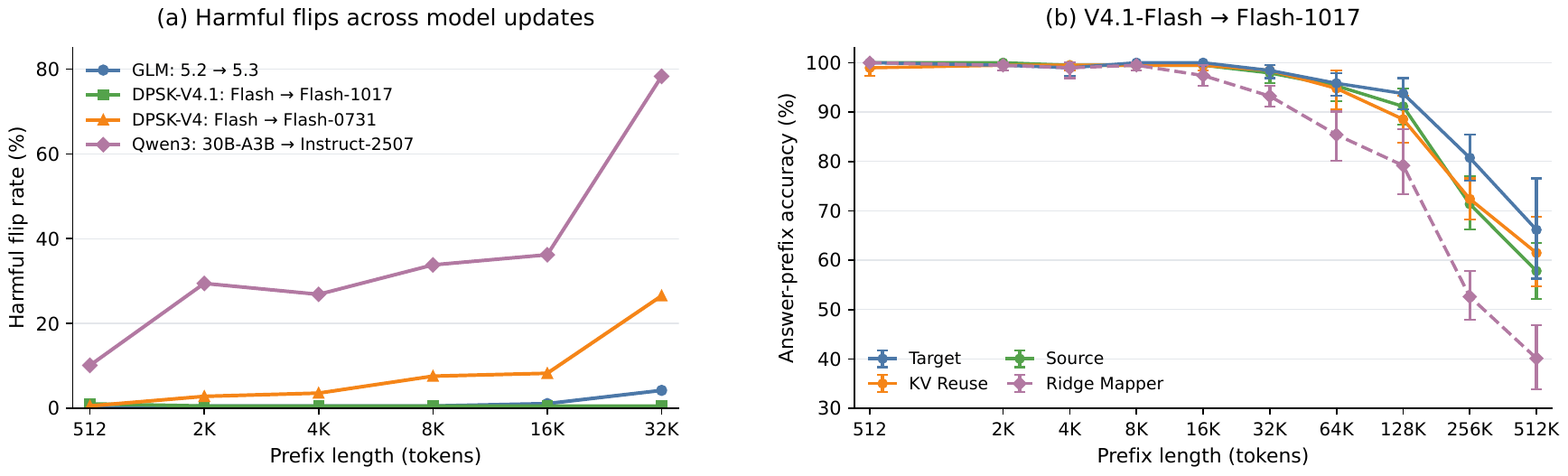}
  \caption{NIAH retrieval under KV reuse. (a) Harmful flip rate in the original sweep across four model updates over 512--32K prefixes, conditional on Target answering correctly. (b) Matched Target, Source, KV Reuse, and Ridge Mapper answer-prefix accuracy for V4.1-Flash $\rightarrow$ Flash-1017 over 512--512K prefixes. Each length has 192 queries from eight histories; error bars in (b) show 95\% history-cluster bootstrap intervals. Data construction and runtime details are given in Appendices~\ref{app:niah} and~\ref{app:runtime}.}
  \label{fig:niah-harmful-flips}
\end{figure}

Figure~\ref{fig:niah-harmful-flips}(a) examines whether reuse preserves answers that Target gets right. We define the \emph{harmful flip rate} (HFR) as $\mathrm{HFR}=100\times N_{\mathrm{flip}}/N_T$. For each model pair and length, $N_T$ counts Target-correct queries, and $N_{\mathrm{flip}}$ counts those answered incorrectly under reuse. Lower HFR means fewer correct answers are lost. At 32K, HFR remains low for GLM (4.22\%) and V4.1-Flash (0.53\%), but rises to 26.53\% for V4-Flash and 78.30\% for Qwen. An overall accuracy gain can still hide losses: V4-Flash reuse breaks 13 of Target's 49 correct answers at 32K but repairs 24 errors.

We extend the V4.1-Flash $\rightarrow$ Flash-1017 evaluation to 512K tokens to test whether its near-native retrieval persists at longer prefixes (Figure~\ref{fig:niah-harmful-flips}(b)). At 64K, KV Reuse remains close to Target, with accuracies of 94.79\% and 95.83\%, respectively. Across 128K--512K, both become less accurate, and reuse falls below Target by 5.21, 8.33, and 4.69 percentage points. At 512K, reuse achieves 61.46\%, compared with 66.15\% for Target and 57.81\% for Source. Thus, reuse retains substantial retrieval ability in these longer settings, but the close agreement observed at shorter prefixes does not persist throughout the extended sweep. The additional loss from reuse does not increase monotonically with length.

Ridge Mapper closely matches direct reuse through 8K, then falls behind as prefixes grow. At 256K and 512K, its accuracies are 52.60\% and 40.10\%, compared with 72.40\% and 61.46\% for direct reuse, gaps of 19.79 and 21.35 percentage points. Direct reuse therefore preserves long-context retrieval more effectively than the calibrated ridge mapping in the V4.1-Flash update.

Our starting point is whether, after a rolling model update, the new model can still interpret and retrieve information from the prefix state computed by its predecessor. GLM and DeepSeek-V4.1 show strong retrieval and reliable reuse over 512--32K prefixes, which may reflect their greater model capacity and stronger retrieval ability. DeepSeek-V4 becomes less accurate at longer prefixes, yet the cache from its stronger source checkpoint still improves the target's retrieval performance, even as the harmful flip rate increases. Qwen offers a complementary contrast: the source scores zero under the tested answer-prefix protocol because of its thinking-style output within the eight-token budget, but the updated target retrieves many values from the same source cache, reaching 80.21\% accuracy at 512 tokens. This shows that a source's answer format and the usefulness of its cached representations can differ. Together, these observations are consistent with a partially shared coordinate system for KV representations across checkpoints of the same family and architecture, with positional alignment where needed. The transferred cache can carry useful information and aspects of the source's prompt-conditioned behavior, while the target's own processing of the query and its readout determine how this state contributes to the answer.

\subsection{Natural-Language QA Evaluation}
\label{sec:quality-latency}

Following the controlled retrieval evaluation, we assess answer quality and acceleration on long-context natural-language QA tasks for the DeepSeek-V4-Flash $\rightarrow$ DeepSeek-V4-Flash-0731 update. On 585 examples, KV Reuse achieves median TTFT speedups of $5.08\times$ on HotpotQA, $2.10\times$ on 2WikiMQA, and $2.51\times$ on MultiFieldQA-en; BABILong reaches $0.86\times$ (Figure~\ref{fig:multibench-quality-ttft}). Target/Reuse scores remain close on BABILong (59.0/58.7 accuracy) and MultiFieldQA-en (45.9/46.2 F1). HotpotQA F1 falls from 69.3 to 66.3, and 2WikiMQA from 61.7 to 54.7: paired changes are $-3.0$ points (95\% CI: $[-6.6,-0.1]$) and $-7.0$ points ($[-12.7,-1.5]$), respectively.

{\centering
  \includegraphics[width=\linewidth]{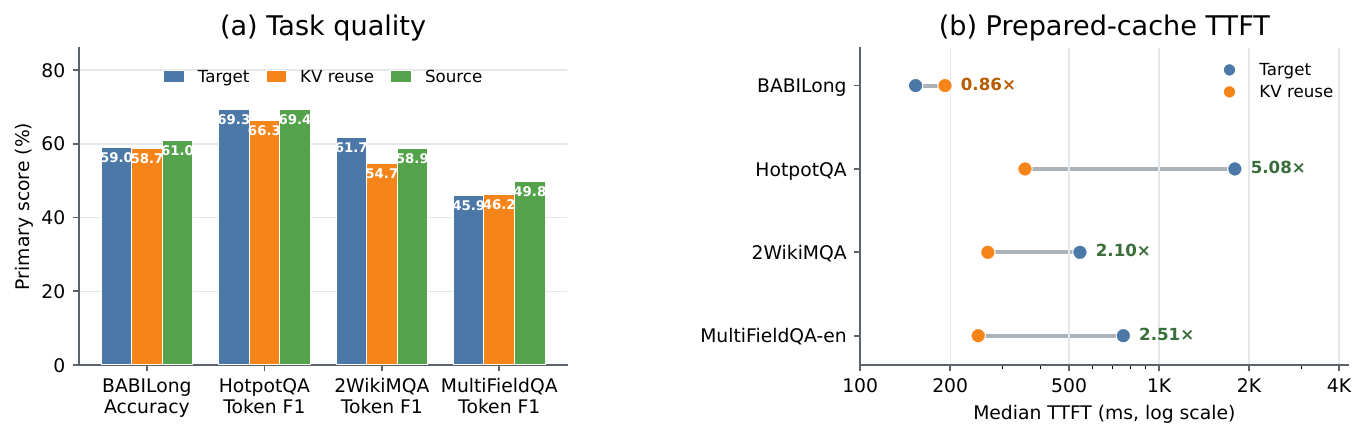}\par
  \captionsetup{hypcap=false}
  \captionof{figure}{Task quality (0--100) and prepared-cache TTFT for Flash $\rightarrow$ Flash-0731. Failed requests score zero. Speedups are median paired Target/Reuse TTFT ratios. TTFT includes cache loading and restoration; source prefill and model startup are excluded.}
  \label{fig:multibench-quality-ttft}
}

Reuse saves more prefill computation on longer prefixes. Median prefixes are 14.3K, 6.4K, and 7.7K tokens for HotpotQA, 2WikiMQA, and MultiFieldQA-en, respectively; BABILong's shorter 1.7K prefix offers less computation to avoid, so cache-access costs outweigh the saving. Quality depends on the task: MultiFieldQA-en retains its score while gaining speed, whereas both multi-hop QA tasks lose F1. The results therefore show substantial acceleration on longer inputs with a task-dependent quality tradeoff. Appendix~\ref{app:results} gives full results and runtime controls.
\par

\subsection{KV Reuse in Earlier and Later Checkpoints}
\label{sec:cache-answer-behavior}
\label{sec:answer-quality}

\begin{figure}[!htbp]
  \centering
  \includegraphics[width=\linewidth]{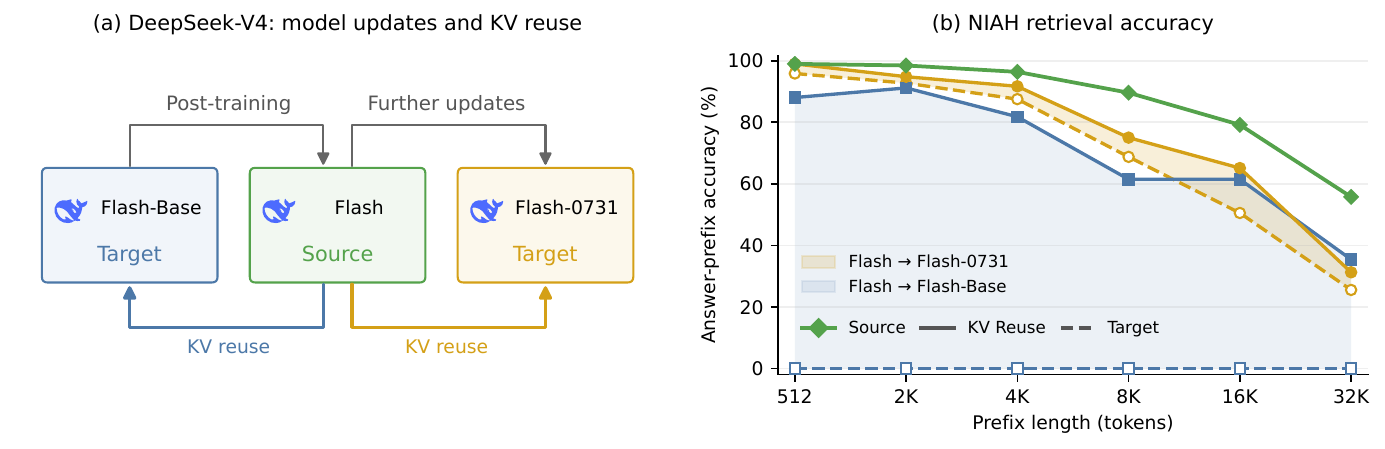}
  \caption{Flash KV reuse in Base and Flash-0731. (a) Model development and cache-transfer directions. (b) NIAH accuracy across 512--32K prefixes. Solid and dashed curves denote KV Reuse and Target. Shading marks their accuracy difference. The green curve shows Flash's source accuracy.}
  \label{fig:niah-forward-base}
\end{figure}

DeepSeek-V4-Flash-Base $\rightarrow$ DeepSeek-V4-Flash $\rightarrow$ DeepSeek-V4-Flash-0731 traces the progression from a pretrained base model through post-training to a later update. We use Flash as the common cache producer and inject its prefix KV states into both Base and Flash-0731 (Figure~\ref{fig:niah-forward-base}(a)). Each receiver uses its own weights to process the question and generate an answer. This compares transfer toward an earlier training stage and a later model version. Figure~\ref{fig:niah-forward-base}(b) shows native and reuse accuracy across 512--32K prefixes, with shading indicating the accuracy change for each receiver.

Flash-produced caches improve retrieval for both receivers. Flash-0731 gains accuracy at every tested length, rising from 50.52\% to 65.10\% at 16K and from 25.52\% to 31.25\% at 32K. The improvement is more pronounced for Base. With its own cache, Base typically restates the question under the instruction-formatted prompt and scores 0\% answer-prefix accuracy. With Flash's KV states, it reaches 88.02\%, 91.15\%, and 81.77\% at 512, 2K, and 4K. Accuracy remains 61.46\% at 16K and 35.42\% at 32K.

These results show that Flash-produced prefix states remain useful at both ends of the model's development. They help Flash-0731 retain part of the source's retrieval performance and help Base return the requested values. The prefix contains both the records and the instruction to return only the value, so the transferred state reflects Flash's processing of content and response instructions together. Injecting these states can therefore change how a receiver uses the context and answers the question while its weights remain fixed. Appendix~\ref{app:base-readout} gives additional Base readout diagnostics.

\subsection{Answer Content and Stopping Under KV Reuse}
\label{sec:sequence-agreement}
\label{sec:types-of-change}

The preceding evaluations examine retrieval, QA quality and speed, and transfer across training stages. We now compare generated answers in the Flash $\rightarrow$ Flash-0731 update, using the separate retrieval cohort and all four benchmarks from Section~\ref{sec:quality-latency}. Table~\ref{tab:generation-and-quality} groups response pairs by whether their recorded token sequences match in full, first differ after the initial token, or differ at the initial token. Sequence comparison includes EOS when present.

\begin{table}[!ht]
  \centering
  \caption{Generation agreement and answer scores for Flash $\rightarrow$ Flash-0731. All agreement rates and scores are percentages. $\Delta$score is KV Reuse minus Target, in percentage points. Retrieval uses the separate diagnostic cohort, and QA uses the main study in Section~\ref{sec:quality-latency}. Agreement uses successful request pairs, and failed requests score zero. Acc. denotes accuracy.}
  \label{tab:generation-and-quality}
  \small
  \setlength{\tabcolsep}{4pt}
  \begin{tabular}{@{}llcccccc@{}}
\toprule
Task & Metric & \shortstack{Full match} & \shortstack{Later change} & \shortstack{First-token change} & \shortstack{Target\\score $\uparrow$} & \shortstack{KV Reuse\\score $\uparrow$} & $\Delta$score \\
\midrule
Retrieval & Acc. & 85.4 & 3.1 & 11.5 & 83.33 & 89.58 & $+6.25$ \\
BABILong & Acc. & 66.3 & 4.7 & 29.0 & 59.00 & 58.67 & $-0.33$ \\
HotpotQA & F1 & 70.0 & 24.0 & 6.0 & 69.28 & 66.28 & $-2.99$ \\
2WikiMQA & F1 & 60.2 & 28.6 & 11.2 & 61.68 & 54.65 & $-7.03$ \\
MultiFieldQA-en & F1 & 41.2 & 46.2 & 12.5 & 45.92 & 46.24 & $+0.32$ \\
\bottomrule
\end{tabular}

\end{table}

Retrieval responses mostly match in full (85.4\%), with only 3.1\% first differing later. In the main QA study, later changes occur in 4.7\% of BABILong responses, 24.0\% of HotpotQA responses, and 28.6\% of 2WikiMQA responses. MultiFieldQA-en has the most frequent later changes (46.2\%), despite 87.5\% first-token agreement. Its mean F1 nevertheless remains close under reuse (45.92\% versus 46.24\%), while 2WikiMQA F1 falls from 61.68\% to 54.65\%. Generation agreement and answer quality therefore capture different effects of reuse. Appendix~\ref{app:main-results} gives the score confidence intervals.

\begin{figure}[!htbp]
  \centering
  \includegraphics[width=0.8\linewidth]{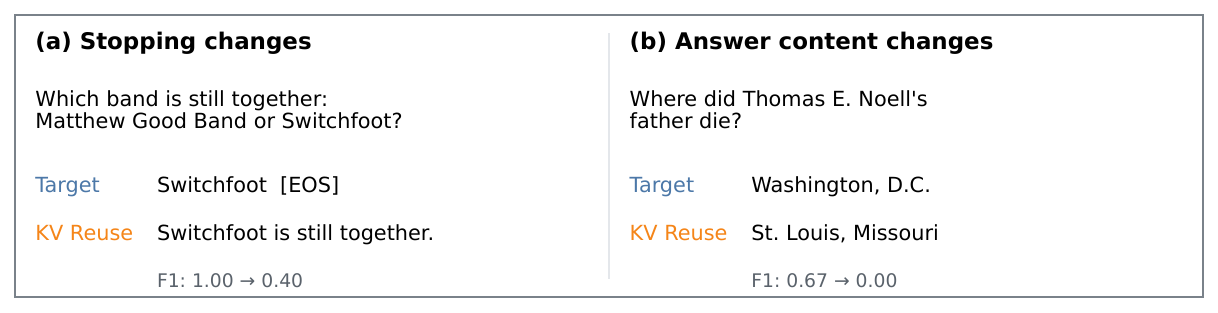}
  \caption{Two diagnostic QA cases illustrate (a) stopping changes and (b) answer-content changes under KV Reuse. Scores are full-response F1.}
  \label{fig:generation-changes}
  \label{fig:output-change-examples}
\end{figure}
\FloatBarrier

Figure~\ref{fig:generation-changes} uses two examples from the detailed QA diagnostics to illustrate how the answers change. For the band question, both runs produce \emph{Switchfoot}. Target then stops, while KV Reuse adds \emph{is still together.} The answer entity is preserved, but the additional wording lowers F1 from 1.00 to 0.40 against the one-word reference. For the question about where Thomas E. Noell's father died, Target answers \emph{Washington, D.C.}, while KV Reuse answers \emph{St. Louis, Missouri}, lowering F1 from 0.67 to 0.00. The context associates Washington with the father and St. Louis with the son, so this case changes the person--location relation used to answer the question.

KV Reuse can therefore affect both the selected answer and the decision to stop after it. These effects help explain why similar first-token predictions can lead to different complete responses and why an F1 loss can reflect either added wording or a changed answer. Section~\ref{sec:explaining-compatibility} examines predictive-distribution changes and how internal-state differences affect retrieval-token decisions.

\section{How KV Reuse Changes Model Predictions}
\label{sec:explaining-compatibility}

This section examines how cache reuse changes the target's computation and predictions. We first compare predictive distributions on retrieval and natural-text continuation for the Flash $\rightarrow$ Flash-0731 update. We then contrast internal-state and answer-margin changes in Flash-0731 and Flash-Base receivers. Finally, controlled interventions in Flash-0731 separate the effects of changed incoming representations and local caches.

\subsection{Functional Divergence Varies Across Tasks}
\label{sec:fcd-tasks}
\label{sec:fcd}

We use \emph{Functional Cache Divergence} (FCD) to measure how much KV reuse changes the target model's next-token predictions. FCD compares the complete probability distributions produced from the target's own prefix cache and the source cache, holding the target model, subsequent input tokens, and prediction position fixed. It reveals changes in the probabilities assigned to candidate tokens even when both runs select the same token, adding information beyond answer accuracy and output agreement.

Let $p_j$ and $q_j$ denote the two distributions at position $j$, with $m_j=(p_j+q_j)/2$. We use Jensen--Shannon divergence with natural logarithms,
\begin{equation}
  \operatorname{FCD}_j
  =\frac{100}{\ln 2}\operatorname{JS}(p_j,q_j)
  =\frac{50}{\ln 2}\left[
    \operatorname{KL}(p_j\Vert m_j)+\operatorname{KL}(q_j\Vert m_j)
  \right].
  \label{eq:fcd-pointwise}
\end{equation}
FCD ranges from 0 (identical distributions) to 100. Larger values indicate greater deviation from target-native predictions, without assigning that change a beneficial or harmful outcome. We average positions within each query or document, queries within each history, and histories or documents equally.

We measure the first NIAH answer position on 192 queries from eight histories per prefix length, and 32 matched continuation positions from each of 64 FineWeb-Edu documents. Both tasks cover 512--32K prefixes. Appendix~\ref{app:fcd-protocol} gives the measurement protocol, and Appendix~\ref{app:fcd-outcomes} reports supplementary statistics.

\Needspace{19\baselineskip}
\begin{wrapfigure}{r}{0.58\textwidth}
  \centering
  \includegraphics[width=\linewidth]{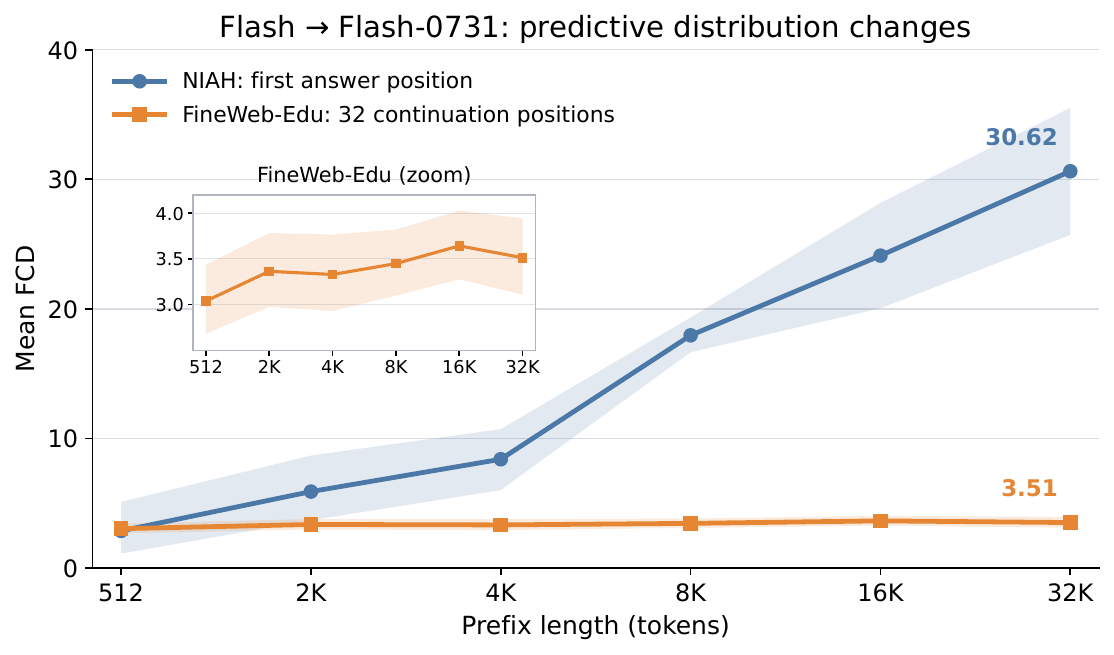}
  \caption{Mean FCD for Flash $\rightarrow$ Flash-0731 on NIAH and FineWeb-Edu. Curves share a vertical scale, with a FineWeb-Edu zoom inset. Bands show cluster-bootstrap 95\% intervals.}
  \label{fig:fcd-tasks}
\end{wrapfigure}

Figure~\ref{fig:fcd-tasks} shows a widening gap between retrieval and natural-text continuation. At 512 tokens, mean FCD is close to 3 for both NIAH (2.87) and FineWeb-Edu (3.04). By 32K, NIAH reaches 30.62, while FineWeb-Edu remains at 3.51. Across the six lengths, FineWeb-Edu stays within 3.04--3.64. The inset shows these small fluctuations, which are modest compared with the rise in NIAH divergence.

The contrast is consistent with how the two tasks use context. NIAH retrieves arbitrary key--value associations, so the answer must come from the cached records. Longer prefixes contain more competing records at the fixed density. Retrieval may therefore be especially sensitive to changes in how the cache represents the queried key and its value. FineWeb-Edu continuation also has cues from nearby words, grammar, and semantic continuity. The target may use these cues to maintain similar predictions after the cache is replaced, offering a plausible explanation for its lower, flatter FCD profile. The NIAH trend also accompanies the higher harmful flip rates at long prefixes in Section~\ref{sec:length-compatibility}.

\subsection{Internal State Changes and Retrieval Decisions}
\label{sec:drift-and-margins}
\label{sec:internal-diagnostics}

\begin{figure}[!t]
  \centering
  \includegraphics[width=0.8\linewidth]{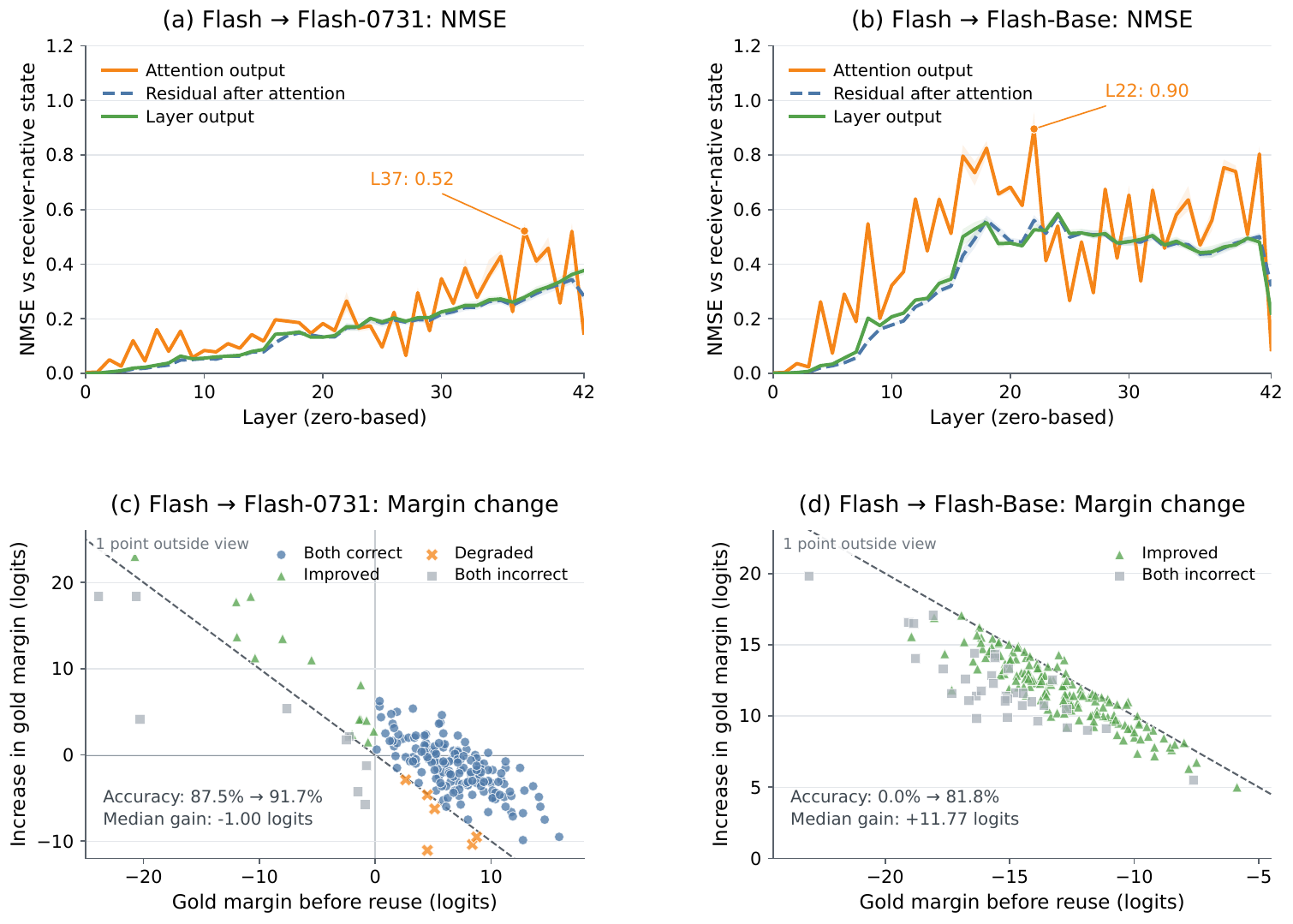}
  \caption{Flash KV reuse in Flash-0731 and Flash-Base. (a, b) Layerwise NMSE over all query positions: medians and 25th--75th percentiles of selected cases. (c, d) Native gold margin versus its change under reuse. Dashed lines mark zero reuse gold margin, and colors show paired answer-prefix correctness. One point per margin panel lies outside the view; all queries enter summaries.}
  \label{fig:drift-and-margins}
\end{figure}

We examine how the internal changes induced by reuse relate to preserved or improved retrieval. Figure~\ref{fig:drift-and-margins} compares Flash-0731 and Flash-Base as receivers of Flash-produced prefix caches. For selected 4K NIAH queries, we trace attention outputs, the residual stream after attention, and layer outputs while the complete query suffix is processed in one forward pass. Within each comparison, receiver weights and input tokens are identical. We measure representation differences with normalized mean squared error (NMSE), and token competition with the gold answer token's score margin over the strongest alternative:
\begin{equation}
  \operatorname{NMSE}(h_R,h_T)
  =\frac{\lVert h_R-h_T\rVert_2^2}{\lVert h_T\rVert_2^2},
  \qquad
  m_{\mathrm{gold}}(z)=z_{\mathrm{gold}}-\max_{v\ne\mathrm{gold}}z_v.
  \label{eq:state-nmse}
\end{equation}
Here, $h_T$ and $h_R$ denote Target and KV Reuse representations at the same layer and site. Margin measurements use the first response-token prediction on all 192 paired queries in the 4K sweep for each receiver. Appendix~\ref{app:diagnostic-data} gives the trace selection and cohort details.

Residual integration often moderates the relative differences introduced by attention. In Figure~\ref{fig:drift-and-margins}(a, b), differences become appreciable in the middle and later layers, with attention-output NMSE fluctuating more sharply than the post-attention residual and layer-output curves. For Flash-0731 at layer 37, median attention-output NMSE is 0.52, compared with 0.27 after residual integration and 0.28 at the layer output. For Flash-Base at layer 22, the corresponding values are 0.90, 0.56, and 0.53. These lower values describe smaller deviations relative to each site's native representation energy. The residual stream moderates this relative disruption at many layers while carrying persistent differences into later computation.

\Needspace{5\baselineskip}
Flash-0731 preserves many correct decisions despite reduced answer margins. Figure~\ref{fig:drift-and-margins}(c) plots $\Delta m=m_{\mathrm{gold}}(z_R)-m_{\mathrm{gold}}(z_T)$ against the native gold margin. Above the dashed threshold ($y=-x$), the gold token has a positive margin under reuse. Many native successes lose some margin but retain enough of their initial lead to remain correct. The median margin change is $-1.00$ logit, yet reuse repairs 14 answers and loses six, raising accuracy from 87.50\% to 91.67\%. Both the initial lead and the direction of the score change help explain which decisions survive.

Flash-Base gains access to cached answers despite substantial state changes. In Figure~\ref{fig:drift-and-margins}(d), every query gains gold margin, with a median increase of 11.77 logits, while answer-prefix accuracy rises from 0 to 81.77\%. Flash-Base often emits whitespace or formatting tokens before the value, so a correct complete answer can still have a negative gold margin at the first response position. The margin shift shows that reuse promotes the answer token, and the complete responses confirm the retrieval gain.

\begin{figure}[!t]
  \centering
  \includegraphics[width=0.85\linewidth]{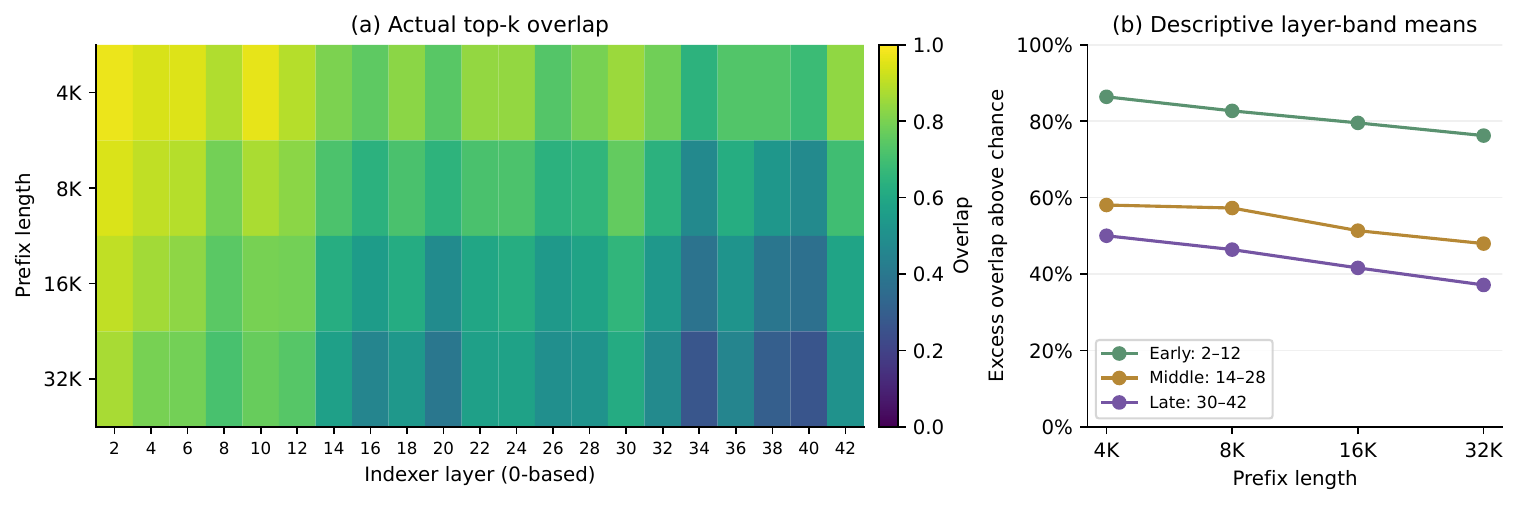}
  \caption{Indexer selection under Flash $\rightarrow$ Flash-Base reuse. (a) Mean top-$k$ overlap with Target across layers. (b) Layer-band means after correcting for uniform-random overlap. Measurements use the first response prediction on four selected retrieval-recovery cases per length. Layer indices are zero-based.}
  \label{fig:indexer-long-context}
\end{figure}

\Needspace{7\baselineskip}
Retrieval recovery also occurs when the receiver accesses a different set of cached entries. The Indexer selects compressed entries for sparse attention. Figure~\ref{fig:indexer-long-context} compares its selections under Target and KV Reuse on four Flash-to-Base recovery cases at each length from 4K to 32K. Mean top-$k$ overlap falls from 81.7\% to 55.4\%, with greater differences in deeper layers. The trend persists after accounting for the overlap expected from uniform random selection. Adjusted overlap decreases from 86.4\% to 76.3\% in early layers and from 50.0\% to 37.1\% in late layers. Yet all these cases answer correctly under reuse after failing with the native cache. Successful retrieval can therefore tolerate substantial changes in which cached entries are selected. Appendix~\ref{app:diagnostic-data} gives the selection metric and sampling details.

Flash-0731 preserves many decisions through sufficient answer margins, while Flash-Base recovers retrieval despite changes in both internal states and Indexer selections. The Base recovery supports the interpretation that useful task state formed by the source remains readable with the receiver's weights. Representation fidelity and retrieval utility are distinct aspects of compatibility.

\subsection{Changes Propagate Across Layers}
\label{sec:input-versus-cache}

We next ask where the internal changes come from, using the separate 96-query Flash-update cohort. Each layer receives the previous layers' computed representation and combines it with its own prefix KV cache. A difference observed at a later layer can therefore arise from that layer's cache or from an input representation already changed by earlier layers. Here, ``input'' means the hidden representation entering the layer, after normalization. The original question tokens remain identical.

We separate these two sources by evaluating target attention as $F(X,C)$ at layers 27, 37, and 42, where $X$ is the incoming representation and $C$ is the prefix cache. Target weights and token positions stay fixed. We compare four combinations: the Target input and cache, the Target input with the source cache, the reuse input with the Target cache, and the reuse input with the source cache. Writing $X_T$ and $X_R$ for the two inputs and $C_T$ and $C_S$ for the two caches,
\begin{equation}
  \begin{aligned}
    A_{TT}&=F(X_T,C_T), & A_{TS}&=F(X_T,C_S),\\
    A_{RT}&=F(X_R,C_T), & A_{RS}&=F(X_R,C_S).
  \end{aligned}
  \label{eq:crossed-attention}
\end{equation}
All output differences use $A_{TT}$ as the reference. Suffix states and index selections are recomputed for each chosen input.

\Needspace{18\baselineskip}
\begin{wrapfigure}{r}{0.50\textwidth}
  \centering
  \includegraphics[width=\linewidth]{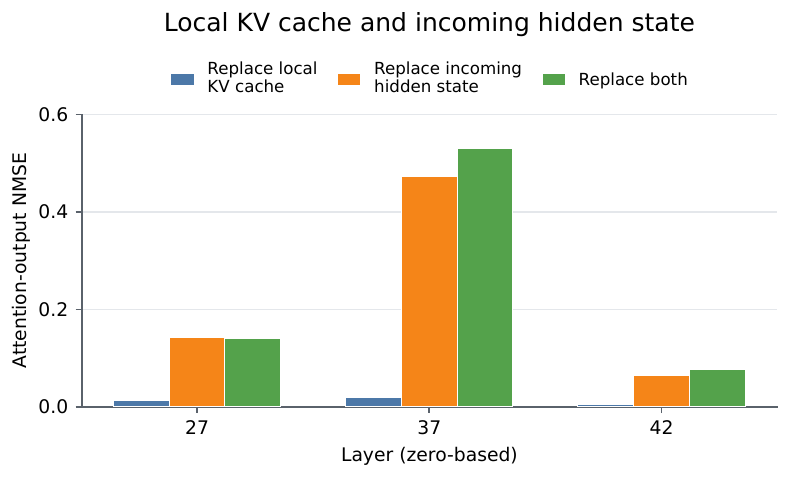}
  \caption{Replacing the local KV cache, the incoming hidden state, or both on the same retrieval cohort. The cache is replaced with the source cache, and the hidden state with the target's representation after earlier layers have reused source caches. Bars show median attention-output NMSE relative to Target. These comparisons do not add up.}
  \label{fig:input-cache-cross}
\end{wrapfigure}

Replacing the incoming representation produces larger changes than replacing only the local cache at all three inspected layers (Figure~\ref{fig:input-cache-cross}). At layer 37, the median output difference is 0.019 when replacing the local KV cache, 0.473 when replacing the incoming hidden state, and 0.530 when both are replaced. Much of the difference at this layer is therefore already present in its incoming representation. The source cache initiates the change, and interactions with earlier layers carry it into later computation. Inspecting a later layer's cache alone misses this propagated effect.

This also helps explain why restoring caches in later layers may not recover the original answer. In four selected retrieval losses, replacing attention outputs in layers 36--42 restores all four answers, whereas replacing those layers' prefix caches restores one. The input representation can still carry earlier changes even after a local cache is restored. These two interventions act at different points and for different durations, so the results illustrate the propagation effect rather than define a cache-repair method. Appendix~\ref{app:crossed} gives the full replacement results.

\section{Related Work}
\label{sec:related-work}

\textbf{KV communication among agents.}
Multi-agent systems can exchange cached representations to avoid repeatedly processing shared context. KVComm~\citep{shi2026kvcomm} selects informative layers' KV pairs for communication. DroidSpeak~\citep{liu2024droidspeak} shares caches between models with the same architecture and selectively recomputes sensitive layers. HeteroFold~\citep{yun2026prefill} extends prefill-free transfer to different model families by aligning model structures, mapping caches, and calibrating receiver behavior. These systems establish KV states as a communication medium. We examine what happens when an updated model directly consumes its predecessor's cache.

\textbf{Cache mapping and alternative reusable states.}
\citet{heo2026cross} fit closed-form ridge maps between selected layers of models in the same family. CacheBridge~\citep{qu2026cachebridge} combines head matching with attention-sensitive calibration to improve transfer quality and reduce mapping cost. RaReCache~\citep{sarkar2026rarecache} selectively recomputes positions where mapped states are unreliable, enabling transfer across larger size gaps. A universal context-reuse layer~\citep{li2026universal} studies translation across differences in scale, architecture, and tokenization. MapFlow~\citep{jin2026accelerating} instead reuses attention maps through translation and partial recomputation. Our main studies analyze direct source-cache transfer; Table~\ref{tab:cross-version-retrieval} additionally compares an adapted closed-form ridge mapper for the Flash update.

\textbf{Prefix reuse and checkpoint compatibility.}
SGLang~\citep{zheng2024sglang} reuses shared-prefix caches, and Prompt Cache~\citep{gim2024prompt} manages reusable prompt modules. CacheBlend~\citep{yao2025cacheblend} uses selective recomputation to combine separately cached chunks. KVShareArena~\citep{shi2026kvsharearena} benchmarks reuse across preceding contexts and model checkpoints, including version-to-version handoffs, and compares quality, computation, memory, and latency. Our study complements this evaluation by holding input tokens fixed and connecting answer quality and generation fidelity to internal-state drift, token competition, and stopping.

\textbf{Compressed caches and positional sensitivity.}
The DeepSeek-V4 report~\citep{deepseekai2026deepseekv4} describes hybrid compressed attention, while DeepSeek-V4.1-Flash~\citep{deepseekai2026deepseekv41flash} further reduces cache storage through compression and sharing. \emph{Periodic Weak Spots}~\citep{zhu2026periodic} shows that retrieval can vary with a token's position relative to compression-window boundaries. This motivates our phase-balanced NIAH queries and reinforces the need to distinguish aggregate retrieval accuracy from position-dependent behavior.

\textbf{Post-training and model iteration.}
LIMA~\citep{zhou2023lima} suggests that instruction tuning can shape response behavior while relying on knowledge acquired during pretraining. The GLM-5 report~\citep{glm5team2026glm5} describes large-scale post-training, and the GLM-5.3 model card~\citep{zai2026glm53} attributes its gains over GLM-5.2 to post-training on the same base model. DeepSeek's Flash-0731 release notes~\citep{deepseekai2026flash0731} likewise specify unchanged architecture and size with renewed post-training. The separate Flash-Base checkpoint~\citep{deepseekai2026flashbase} provides our instruction-training contrast. These releases motivate studying cached knowledge and prompt-conditioned answer behavior together.

\textbf{Long-context evaluation.}
BABILong~\citep{kuratov2024babilong} embeds reasoning tasks in long contexts. LongBench~\citep{bai2024longbench} combines multiple long-context tasks, including MultiFieldQA, while HotpotQA~\citep{yang2018hotpotqa} and 2WikiMQA~\citep{ho2020constructing} test multi-hop question answering. LongBench v2~\citep{bai2024longbenchv2} extends evaluation toward realistic, demanding contexts, and Multi-News~\citep{fabbri2019multinews} supplies multi-document summarization. FineWeb-Edu~\citep{penedo2024fineweb} provides the natural documents used in our matched-continuation FCD evaluation. Our main comparisons use BABILong and LongBench v1 tasks alongside controlled retrieval.

\textbf{Representation comparison and decision analysis.}
Centered kernel alignment~\citep{kornblith2019similarity} compares representations under chosen invariances. Our NMSE measures differences encountered in paired target executions, and logit margins relate those differences to token selection. Causal tracing~\citep{meng2022locating} and activation-patching studies~\citep{zhang2024patching} motivate our interventions on incoming layer inputs, prefix caches, and attention outputs. Comparing token scores at the first retrieval-answer position connects internal differences to changes in the selected token.

\textbf{Inference implementation.}
Transformers~\citep{wolf2020transformers} and PyTorch~\citep{paszke2019pytorch} support the detailed activation and generation analyses, while the serving evaluation uses SGLang. These execution paths have different cache-restoration and generation controls. We therefore report their settings separately and use same-version restoration to measure changes introduced by the handoff itself.

\section{Conclusion and Discussion}
\label{sec:conclusion}

We studied whether DeepSeek-V4-Flash-0731 can continue from a cache produced by DeepSeek-V4-Flash. Reuse retains many retrieval answers despite substantial differences in attention outputs and final representations. With the source cache prepared, it reduces time to first token on the tested benchmarks with longer prefixes, while task scores vary by benchmark. Natural question answering shows why answer quality alone is incomplete: the first token can agree with target-native inference even when later content or stopping changes. The Flash-to-Base retrieval contrast further shows that a cache carries prompt-conditioned state: it contains both the records and the instruction governing how to answer. Across these experiments, task quality, generation fidelity, and state similarity answer different questions about compatibility.

The internal analysis connects persistent state drift to decisions through token-score gaps. Many answers survive because the native winner retains enough advantage over competitors; a smaller shift can flip an answer when its original gap is narrow. At the three inspected layers, changes arriving through the input perturb attention more than changing only the local prefix cache. Earlier interactions with reused caches can create that changed input, so drift accumulates along the target's continuation rather than remaining confined to a single layer's stored state.

Across the four retrieval updates, GLM and V4.1-Flash preserve correct answers most reliably, V4-Flash gains aggregate accuracy, and Qwen loses accuracy under reuse. The V4.1-Flash extension retains 61.46\% accuracy at 512K. Functional divergence reveals a further task contrast: prediction changes grow with prefix length on retrieval and stay low and stable on natural-text continuation. Future work can extend these comparisons to more tasks and decoding settings, predict changes that affect token decisions before reuse, and separate the contributions of cached facts and response instructions.

\clearpage
\bibliographystyle{plainnat}
\bibliography{references}

@article{liu2024droidspeak,
  title   = {DroidSpeak: {KV} Cache Sharing for Cross-{LLM} Communication and Multi-{LLM} Serving},
  author  = {Liu, Yuhan and Huang, Yuyang and Yao, Jiayi and Feng, Shaoting and Gu, Zhuohan and Du, Kuntai and Li, Hanchen and Cheng, Yihua and Jiang, Junchen and Lu, Shan and Musuvathi, Madan and Choukse, Esha},
  journal = {arXiv preprint arXiv:2411.02820},
  year    = {2024},
  url     = {https://arxiv.org/abs/2411.02820}
}

@article{heo2026cross,
  title={Cross-Model {KV} Cache Transfer in {LLM} Families: A Closed-Form Linear Mapping for Prefill Reuse},
  author={Heo, Taekyung and Shafipour, Rasoul and Zhao, Ritchie and Golub, Maximilian and Kamani, Mohammad Mahdi and Borkar, Ritika and Chandran, Makesh Tarun and Zardoshti, Pantea and Rouhani, Bita Darvish},
  journal={arXiv preprint arXiv:2608.03893},
  year={2026},
  url={https://arxiv.org/abs/2608.03893}
}

@misc{qu2026cachebridge,
  title         = {{CacheBridge}: Efficient Cross-Model {KV} Cache Transfer},
  author        = {Qu, Xingyu and Lu, Siyuan and Chen, Zhiyu and Wang, Sheng and Lin, Tao},
  year          = {2026},
  eprint        = {2609.00891},
  archivePrefix = {arXiv},
  primaryClass  = {cs.LG},
  url           = {https://arxiv.org/abs/2609.00891}
}

@article{zheng2024sglang,
  title={Sglang: Efficient execution of structured language model programs},
  author={Zheng, Lianmin and Yin, Liangsheng and Xie, Zhiqiang and Sun, Chuyue and Huang, Jeff and Yu, Cody H and Cao, Shiyi and Kozyrakis, Christos and Stoica, Ion and Gonzalez, Joseph E and others},
  journal={Advances in neural information processing systems},
  volume={37},
  pages={62557--62583},
  year={2024}
}

@article{gim2024prompt,
  title={Prompt cache: Modular attention reuse for low-latency inference},
  author={Gim, In and Chen, Guojun and Lee, Seung-seob and Sarda, Nikhil and Khandelwal, Anurag and Zhong, Lin},
  journal={Proceedings of Machine Learning and Systems},
  volume={6},
  pages={325--338},
  year={2024}
}

@inproceedings{yao2025cacheblend,
  title={Cacheblend: Fast large language model serving for rag with cached knowledge fusion},
  author={Yao, Jiayi and Li, Hanchen and Liu, Yuhan and Ray, Siddhant and Cheng, Yihua and Zhang, Qizheng and Du, Kuntai and Lu, Shan and Jiang, Junchen},
  booktitle={Proceedings of the twentieth European conference on computer systems},
  pages={94--109},
  year={2025}
}

@article{zhou2023lima,
  title={Lima: Less is more for alignment},
  author={Zhou, Chunting and Liu, Pengfei and Xu, Puxin and Iyer, Srinivasan and Sun, Jiao and Mao, Yuning and Ma, Xuezhe and Efrat, Avia and Yu, Ping and Yu, Lili and others},
  journal={Advances in Neural Information Processing Systems},
  volume={36},
  pages={55006--55021},
  year={2023}
}

@inproceedings{kornblith2019similarity,
  title     = {Similarity of Neural Network Representations Revisited},
  author    = {Kornblith, Simon and Norouzi, Mohammad and Lee, Honglak and Hinton, Geoffrey},
  booktitle = {Proceedings of the 36th International Conference on Machine Learning},
  series    = {Proceedings of Machine Learning Research},
  volume    = {97},
  pages     = {3519--3529},
  publisher = {PMLR},
  year      = {2019},
  url       = {https://proceedings.mlr.press/v97/kornblith19a.html}
}

@inproceedings{meng2022locating,
  title     = {Locating and Editing Factual Associations in {GPT}},
  author    = {Meng, Kevin and Bau, David and Andonian, Alex and Belinkov, Yonatan},
  booktitle = {Advances in Neural Information Processing Systems},
  volume    = {35},
  pages     = {17359--17372},
  year      = {2022},
  url       = {https://arxiv.org/abs/2202.05262}
}

@inproceedings{zhang2024patching,
  title     = {Towards Best Practices of Activation Patching in Language Models: Metrics and Methods},
  author    = {Zhang, Fred and Nanda, Neel},
  booktitle = {The Twelfth International Conference on Learning Representations},
  year      = {2024},
  url       = {https://arxiv.org/abs/2309.16042}
}

@misc{glm5team2026glm5,
  title         = {{GLM-5}: from Vibe Coding to Agentic Engineering},
  author        = {{GLM-5-Team}},
  year          = {2026},
  eprint        = {2602.15763},
  archivePrefix = {arXiv},
  primaryClass  = {cs.LG},
  url           = {https://arxiv.org/abs/2602.15763}
}

@misc{zai2026glm53,
  title        = {{GLM-5.3}},
  author       = {{Z.ai}},
  year         = {2026},
  howpublished = {Official model card},
  note         = {Describes the shared base model with GLM-5.2 and improvements from post-training. Accessed October 26, 2026},
  url          = {https://huggingface.co/zai-org/GLM-5.3}
}

@misc{deepseekai2026deepseekv4,
  title         = {{DeepSeek-V4}: Towards Highly Efficient Million-Token Context Intelligence},
  author        = {{DeepSeek-AI}},
  year          = {2026},
  eprint        = {2606.19348},
  archivePrefix = {arXiv},
  primaryClass  = {cs.CL},
  url           = {https://arxiv.org/abs/2606.19348}
}

@misc{deepseekai2026deepseekv41flash,
  title        = {{DeepSeek-V4.1-Flash}: Pushing the Limits of {KV} Cache Compression},
  author       = {{DeepSeek-AI}},
  year         = {2026},
  howpublished = {Technical report},
  url          = {https://huggingface.co/deepseek-ai/DeepSeek-V4.1-Flash/blob/main/DeepSeek_V41_Tech_Report.pdf}
}

@misc{deepseekai2026flash0731,
  title        = {{DeepSeek-V4-Flash} Update},
  author       = {{DeepSeek-AI}},
  year         = {2026},
  month        = jul,
  howpublished = {DeepSeek API Docs, Change Log},
  note         = {Release entries dated April 24 and July 31, 2026. Accessed October 26, 2026},
  url          = {https://api-docs.deepseek.com/updates#date-2026-07-31}
}

@misc{deepseekai2026flashbase,
  title        = {{DeepSeek-V4-Flash-Base}},
  author       = {{DeepSeek-AI}},
  year         = {2026},
  howpublished = {Hugging Face model repository},
  note         = {Accessed October 23, 2026},
  url          = {https://huggingface.co/deepseek-ai/DeepSeek-V4-Flash-Base}
}

@article{penedo2024fineweb,
  title   = {The {FineWeb} Datasets: Decanting the Web for the Finest Text Data at Scale},
  author  = {Penedo, Guilherme and Kydl{\'i}{\v c}ek, Hynek and Ben Allal, Loubna and Lozhkov, Anton and Mitchell, Margaret and Raffel, Colin and von Werra, Leandro and Wolf, Thomas},
  journal = {arXiv preprint arXiv:2406.17557},
  year    = {2024},
  url     = {https://arxiv.org/abs/2406.17557}
}

@inproceedings{kuratov2024babilong,
  title     = {{BABILong}: Testing the Limits of {LLMs} with Long Context Reasoning-in-a-Haystack},
  author    = {Kuratov, Yuri and Bulatov, Aydar and Anokhin, Petr and Rodkin, Ivan and Sorokin, Dmitry and Sorokin, Artyom and Burtsev, Mikhail},
  booktitle = {Advances in Neural Information Processing Systems},
  volume    = {37},
  year      = {2024},
  url       = {https://arxiv.org/abs/2406.10149}
}

@inproceedings{bai2024longbench,
  title     = {{LongBench}: A Bilingual, Multitask Benchmark for Long Context Understanding},
  author    = {Bai, Yushi and Lv, Xin and Zhang, Jiajie and Lyu, Hongchang and Tang, Jiankai and Huang, Zhidian and Du, Zhengxiao and Liu, Xiao and Zeng, Aohan and Hou, Lei and Dong, Yuxiao and Tang, Jie and Li, Juanzi},
  booktitle = {Proceedings of the 62nd Annual Meeting of the Association for Computational Linguistics (Volume 1: Long Papers)},
  pages     = {3119--3137},
  publisher = {Association for Computational Linguistics},
  year      = {2024},
  doi       = {10.18653/v1/2024.acl-long.172},
  url       = {https://aclanthology.org/2024.acl-long.172/}
}

@inproceedings{yang2018hotpotqa,
  title     = {{HotpotQA}: A Dataset for Diverse, Explainable Multi-hop Question Answering},
  author    = {Yang, Zhilin and Qi, Peng and Zhang, Saizheng and Bengio, Yoshua and Cohen, William and Salakhutdinov, Ruslan and Manning, Christopher D.},
  booktitle = {Proceedings of the 2018 Conference on Empirical Methods in Natural Language Processing},
  pages     = {2369--2380},
  publisher = {Association for Computational Linguistics},
  year      = {2018},
  doi       = {10.18653/v1/D18-1259},
  url       = {https://aclanthology.org/D18-1259/}
}

@inproceedings{ho2020constructing,
  title     = {Constructing A Multi-hop {QA} Dataset for Comprehensive Evaluation of Reasoning Steps},
  author    = {Ho, Xanh and Nguyen, Anh-Khoa Duong and Sugawara, Saku and Aizawa, Akiko},
  booktitle = {Proceedings of the 28th International Conference on Computational Linguistics},
  pages     = {6609--6625},
  publisher = {International Committee on Computational Linguistics},
  year      = {2020},
  doi       = {10.18653/v1/2020.coling-main.580},
  url       = {https://aclanthology.org/2020.coling-main.580/}
}

@inproceedings{wolf2020transformers,
  title     = {{Transformers}: State-of-the-Art Natural Language Processing},
  author    = {Wolf, Thomas and Debut, Lysandre and Sanh, Victor and Chaumond, Julien and Delangue, Clement and Moi, Anthony and Cistac, Pierric and Rault, Tim and Louf, Remi and Funtowicz, Morgan and Davison, Joe and Shleifer, Sam and von Platen, Patrick and Ma, Clara and Jernite, Yacine and Plu, Julien and Xu, Canwen and Le Scao, Teven and Gugger, Sylvain and Drame, Mariama and Lhoest, Quentin and Rush, Alexander},
  booktitle = {Proceedings of the 2020 Conference on Empirical Methods in Natural Language Processing: System Demonstrations},
  pages     = {38--45},
  publisher = {Association for Computational Linguistics},
  year      = {2020},
  doi       = {10.18653/v1/2020.emnlp-demos.6},
  url       = {https://aclanthology.org/2020.emnlp-demos.6/}
}

@inproceedings{paszke2019pytorch,
  title     = {{PyTorch}: An Imperative Style, High-Performance Deep Learning Library},
  author    = {Paszke, Adam and Gross, Sam and Massa, Francisco and Lerer, Adam and Bradbury, James and Chanan, Gregory and Killeen, Trevor and Lin, Zeming and Gimelshein, Natalia and Antiga, Luca and Desmaison, Alban and K{\"o}pf, Andreas and Yang, Edward and DeVito, Zach and Raison, Martin and Tejani, Alykhan and Chilamkurthy, Sasank and Steiner, Benoit and Fang, Lu and Bai, Junjie and Chintala, Soumith},
  booktitle = {Advances in Neural Information Processing Systems},
  volume    = {32},
  year      = {2019},
  url       = {https://arxiv.org/abs/1912.01703}
}

@inproceedings{fabbri2019multinews,
  title     = {{Multi-News}: A Large-Scale Multi-Document Summarization Dataset and Abstractive Hierarchical Model},
  author    = {Fabbri, Alexander and Li, Irene and She, Tianwei and Li, Suyi and Radev, Dragomir},
  booktitle = {Proceedings of the 57th Annual Meeting of the Association for Computational Linguistics},
  pages     = {1074--1084},
  publisher = {Association for Computational Linguistics},
  year      = {2019},
  doi       = {10.18653/v1/P19-1102},
  url       = {https://aclanthology.org/P19-1102/}
}

@article{bai2024longbenchv2,
  title   = {{LongBench v2}: Towards Deeper Understanding and Reasoning on Realistic Long-context Multitasks},
  author  = {Bai, Yushi and Tu, Shangqing and Zhang, Jiajie and Peng, Hao and Wang, Xiaozhi and Lv, Xin and Cao, Shulin and Xu, Jiazheng and Hou, Lei and Dong, Yuxiao and Tang, Jie and Li, Juanzi},
  journal = {arXiv preprint arXiv:2412.15204},
  year    = {2024},
  url     = {https://arxiv.org/abs/2412.15204}
}

@inproceedings{shi2026kvcomm,
  title={{KVComm}: Enabling Efficient {LLM} Communication through Selective {KV} Sharing},
  author={Shi, Xiangyu and Chiesa, Marco and Maguire, Jr., Gerald Q. and Kostic, Dejan},
  booktitle={International Conference on Learning Representations},
  year={2026},
  url={https://arxiv.org/abs/2510.03346}
}

@article{li2026universal,
  title={A Universal Context-Reuse Layer for Cross-Model KV Sharing},
  author={Li, Yi and Jiang, Dongming and Zhao, Yi and Li, Bingzhe},
  journal={arXiv preprint arXiv:2608.30963},
  year={2026},
  url={https://arxiv.org/abs/2608.30963}
}

@article{yun2026prefill,
  title={Prefill-Free Cross-Family KV Cache Transfer for Heterogeneous Multi-Agent LLMs},
  author={Yun, Vincent-Daniel and Lim, Woosang and Yoo, Haneul and Yoo, Sungjoo and Annavaram, Murali and Karimireddy, Sai Praneeth},
  journal={arXiv preprint arXiv:2609.32259},
  year={2026},
  url={https://arxiv.org/abs/2609.32259}
}

@article{zhu2026periodic,
  title={Periodic Weak Spots: Phase Sensitivity from Chunked KV-Cache Compression},
  author={Zhu, Xingyu and Yi, Pu and Cheng, Ziheng and Lv, Ang and Liu, Jing and Ying, Lexing and Ma, Yiyuan and Dong, Xin},
  journal={arXiv preprint arXiv:2609.36322},
  year={2026},
  url={https://arxiv.org/abs/2609.36322}
}

@article{sarkar2026rarecache,
  title={RaReCache: Bridging the Gap in Cross-Model KV Cache Reuse via Rank disagreement-based Selective Recomputation},
  author={Sarkar, Sreetama and Mitra, Saptarshi and Huang, Sitao and Kundu, Souvik and Beerel, Peter A},
  journal={arXiv preprint arXiv:2610.11358},
  year={2026},
  url={https://arxiv.org/abs/2610.11358}
}

@article{shi2026kvsharearena,
  title={KVShareArena: KV-Cache Reuse Across Contexts and Model Checkpoints},
  author={Shi, Xi and Zheng, Mengxin and Lou, Qian},
  journal={arXiv preprint arXiv:2609.10266},
  year={2026},
  url={https://arxiv.org/abs/2609.10266}
}

@article{jin2026accelerating,
  title={Accelerating Cross-Model LLM Inference via Attention Map Reuse},
  author={Jin, Zhan and Luo, Yuze and Cui, Bin and Miao, Xupeng},
  journal={arXiv preprint arXiv:2610.14688},
  year={2026},
  url={https://arxiv.org/abs/2610.14688}
}

@article{hsieh2024ruler,
  title={RULER: What's the real context size of your long-context language models?},
  author={Hsieh, Cheng-Ping and Sun, Simeng and Kriman, Samuel and Acharya, Shantanu and Rekesh, Dima and Jia, Fei and Zhang, Yang and Ginsburg, Boris},
  journal={arXiv preprint arXiv:2404.06654},
  year={2024}
}

\clearpage
\appendix

\section*{Appendix Contents}
\vspace{1em}
\begin{tabular*}{\textwidth}{@{\extracolsep{\fill}}p{0.86\textwidth}r@{}}
  \hyperref[app:data]{\textbf{A}\quad Tasks, Datasets, and Evaluation Protocols} & \pageref{app:data} \\
  \hspace*{1.5em}\hyperref[app:niah]{A.1\quad Retrieval Compatibility} & \pageref{app:niah} \\
  \hspace*{1.5em}\hyperref[app:datasets]{A.2\quad Task Quality and Latency} & \pageref{app:datasets} \\
  \hspace*{1.5em}\hyperref[app:diagnostic-data]{A.3\quad Behavioral and Mechanistic Analysis} & \pageref{app:diagnostic-data} \\
  \hspace*{1.5em}\hyperref[app:fcd-protocol]{A.4\quad Functional Cache Divergence} & \pageref{app:fcd-protocol} \\
  \addlinespace
  \hyperref[app:implementation]{\textbf{B}\quad Cache Handoff and Runtime Controls} & \pageref{app:implementation} \\
  \hspace*{1.5em}\hyperref[app:cache]{B.1\quad Composite Cache and Page Validation} & \pageref{app:cache} \\
  \hspace*{1.5em}\hyperref[app:runtime]{B.2\quad Runtime Matrix} & \pageref{app:runtime} \\
  \addlinespace
  \hyperref[app:results]{\textbf{C}\quad Extended Results} & \pageref{app:results} \\
  \hspace*{1.5em}\hyperref[app:main-results]{C.1\quad Complete Benchmark Results} & \pageref{app:main-results} \\
  \hspace*{1.5em}\hyperref[app:length-results]{C.2\quad Length-Stratified Results} & \pageref{app:length-results} \\
  \hspace*{1.5em}\hyperref[app:controls]{C.3\quad Same-Version Controls and Sensitivity Checks} & \pageref{app:controls} \\
  \hspace*{1.5em}\hyperref[app:output-review]{C.4\quad Output Divergence} & \pageref{app:output-review} \\
  \hspace*{1.5em}\hyperref[app:fcd-outcomes]{C.5\quad Functional Divergence by Retrieval Outcome} & \pageref{app:fcd-outcomes} \\
  \addlinespace
  \hyperref[app:mechanism]{\textbf{D}\quad NIAH Readout and Mechanism Details} & \pageref{app:mechanism} \\
  \hspace*{1.5em}\hyperref[app:base-readout]{D.1\quad Flash-to-Base Readout Diagnostics} & \pageref{app:base-readout} \\
  \hspace*{1.5em}\hyperref[app:crossed]{D.2\quad Crossed Input--Cache Intervention} & \pageref{app:crossed} \\
  \hspace*{1.5em}\hyperref[app:perturbation]{D.3\quad Matched Perturbations and Stopping} & \pageref{app:perturbation} \\
\end{tabular*}
\clearpage

\section{Tasks, Datasets, and Evaluation Protocols}
\label{app:data}

This appendix follows the four evaluations in Section~\ref{sec:tasks-and-datasets}, describing their data construction, sample selection, and evaluation settings. Prefix lengths count the tokenized history and exclude the question or continuation suffix. Runtime and cache-handoff controls are given in Appendix~\ref{app:implementation}.

\subsection{Retrieval Compatibility}
\label{app:niah}

The NIAH sweeps use eight independent histories at each of six prefix lengths: 512, 2K, 4K, 8K, 16K, and 32K tokens. Each history supplies 24 queries targeting records near 10\%, 50\%, and 90\% of the prefix, balanced across eight placement phases. This yields 192 queries per length and 1,152 per model pair. Supplementary near-boundary queries are excluded from the reported retrieval results.

Records are interleaved with fixed filler text at approximately one record per 16 prefix tokens. The 512-, 2K-, and 4K-token histories contain 32, 128, and 256 records, respectively. Every key and value is checked to encode as a single token, remain unchanged at the tokenizer boundary in the \texttt{key:value} record, and remain distinct from the instruction and filler vocabulary. Construction uses each model pair's tokenizer. Paired configurations receive identical token IDs.

For Flash, the prefix begins with ``Read these records carefully. Every key has exactly one value. Use the records to answer the question. Return only the value.'' The query suffix is ``What is the value for key:KEY? Answer with only the value,'' followed by the official assistant-generation prefix. GLM uses a plain completion prompt ending in ``Question: What is the value for key:KEY? Answer:''. Pro, V4.1, and Qwen use their model-specific prompt encodings. Table~\ref{tab:app-niah-example} shows one actual 512-token Flash example.

\begin{table}[!htbp]
  \centering
  \caption{A Flash NIAH example. Prefix text is abbreviated; chat-template tokens and record separators are omitted for readability.}
  \label{tab:app-niah-example}
  \small
  \renewcommand{\arraystretch}{1.15}
  \begin{tabularx}{\textwidth}{@{}>{\raggedright\arraybackslash}p{0.16\textwidth}>{\raggedright\arraybackslash}X@{}}
    \toprule
    Component & Example \\
    \midrule
    Prefix $P$ (excerpt) &
      \begin{minipage}[t]{\linewidth}\raggedright\ttfamily
        Read these records carefully. Every key has exactly one value. Use the records to answer the question. Return only the value.
        \par\smallskip
        The morning light fell across the quiet garden.
        \par\smallskip
        \ldots{} \textbf{yet:vector} \ldots{} ynt:izio \ldots{} IELD:Popular \ldots{}
      \end{minipage} \\
    \midrule
    Query $Q$ & \texttt{What is the value for key:yet? Answer with only the value.} \\
    \midrule
    Gold answer & \texttt{vector} \\
    \bottomrule
  \end{tabularx}
\end{table}

All retrieval runs use greedy decoding, with a maximum of eight output tokens for GLM, V4-Flash, and Qwen, and one for Pro and V4.1-Flash, including the long-context extension. Answer-prefix accuracy counts a response as correct when its whitespace-trimmed text begins with the requested value. Failed requests score zero. For Qwen, Source's thinking-format output can fill the eight-token budget before an answer is returned. The zero Source score therefore describes the tested response protocol.

The Flash-update and Flash-to-Base short-prefix analyses use the 512-, 2K-, and 4K-token subset of the six-length Flash sweep, comprising 576 queries from 24 histories. The Pro evaluation uses the same length and sample-count design. Appendix~\ref{app:base-readout} reports the Flash-to-Base readout statistics.

The V4.1-Flash extension adds 64K, 128K, 256K, and 512K prefixes using the same history count, record density, and depth-query design. Its 768 added queries bring the full sweep in Figure~\ref{fig:niah-harmful-flips}(b) to 1,920 queries. Keys and values remain globally disjoint through 256K. At 512K, the required 65,536 distinct terms exceed the available 53,162-term pool, so a token may serve as a key in one record and a value in another. Keys remain unique, values remain unique, and no record binds a key to itself. Runtime differences between the short and long sweeps are described in Appendix~\ref{app:runtime}.

\paragraph{Closed-form mapper comparison.}
\label{app:mapper-protocol}
The Flash block in Table~\ref{tab:cross-version-retrieval} evaluates an adaptation of the centered ridge mapping and source-layer selection of \citet{heo2026cross}, rather than an author-provided implementation. We calibrate on 500 independent FineWeb-Edu documents, using 1,024 prefix tokens per document with both models. A separate set of 32 documents supplies 16 matched continuation positions per document to choose $k\in\{1,2,4\}$ by the lowest mean full-vocabulary FCD; this selects $k=4$. These documents are disjoint from the report's earlier natural-text evaluation sets, and no NIAH answers are used for fitting or selection.

For each primary target-cache component, source layers are ranked by ridge prediction $R^2$ within compatible component types, widths, and compression rates. The selected $k$ source vectors are concatenated into $X$, paired with target vectors $Y$, and centered to obtain $X_c$ and $Y_c$. We fit the full affine map
\begin{equation}
  W=(X_c^\top X_c+\lambda I)^{-1}X_c^\top Y_c,
  \qquad M(x)=(x-\bar{x})W+\bar{y},
  \label{eq:mapper-ridge}
\end{equation}
using $\lambda=0.01$ for the summed squared-error objective and streamed float64 sufficient statistics, without a diagonal approximation. Primary uncompressed KV rows are sampled every four tokens and auxiliary projection rows every 16; source-layer ranking uses at most 2,048 paired rows per component.

DeepSeek's shared K/V vectors are mapped jointly in content coordinates before the trailing 64-dimensional interleaved RoPE rotation; target runtime frequencies, scales, and absolute positions are then applied. Mapping covers the active window, compressed, and indexer KV states throughout the target's 43 layers, together with same-layer full-ridge maps for nonempty compressor/indexer buffer, gate, and overlap states. Cache counters and positions are preserved, and K/V sharing is restored. This complete-state adaptation is checked by pre-RoPE capture validation and identity transfers with exact cache and logit agreement at 512 and 1,031 tokens, including nonempty compression tails.

All four configurations are rerun on the same 1,152 fixed NIAH queries with identical token IDs, prefill chunks, and eight-token greedy decoding. The Flash baselines in Table~\ref{tab:cross-version-retrieval} come from this matched run; earlier figures and diagnostic analyses retain their original measurement campaigns. Accuracy differences use paired history-cluster bootstrap intervals with 10,000 replicates. The mapper comparison assesses retrieval correctness in the Transformers path and does not provide a serving-latency comparison.

For V4.1-Flash $\rightarrow$ Flash-1017, we fit a separate centered ridge mapper on 500 FineWeb-Edu documents with 1,024-token prefixes. Held-out FCD on 32 documents with 16 continuation positions each selects $k=4$ from $\{1,2,4\}$, capped by the number of compatible source cache producers for each component. The fit uses $\lambda=0.01$ and normalized pre-RoPE cache vectors. Mapped states are re-encoded in the native cache formats, compressor and indexer tails are mapped separately, and Engram token histories are preserved.

Figure~\ref{fig:niah-harmful-flips}(b) uses a matched rerun of all four configurations on the same 1,920 fixed queries over 512--512K prefixes, with 8,192-token prefill chunks and one-token greedy decoding. Its intervals use 10,000 paired history-cluster bootstrap draws. Figure~\ref{fig:cross-version-overview} and Figure~\ref{fig:niah-harmful-flips}(a) retain the original sweep, while panel (b) uses the baselines measured alongside Ridge Mapper.

\FloatBarrier
\subsection{Task Quality and Latency}
\label{app:datasets}

The 585-sample study evaluates Flash $\rightarrow$ Flash-0731 on fixed snapshots of BABILong~\citep{kuratov2024babilong} and LongBench v1~\citep{bai2024longbench}. Samples are selected independently of model outcomes. Table~\ref{tab:app-datasets} gives the source snapshots, eligible pools, and evaluated sample counts.

\begin{table}[!htbp]
  \centering
  \caption{Dataset provenance and main-study selection. ``Eligible'' counts rows surviving input and tokenizer checks; ``independent'' counts distinct stories or source documents. BABILong contains 20 examples for every QA-type and length pair.}
  \label{tab:app-datasets}
  \footnotesize
  \setlength{\tabcolsep}{3pt}
  \begin{tabular}{@{}>{\raggedright\arraybackslash}p{0.17\textwidth}>{\raggedright\arraybackslash}p{0.24\textwidth}>{\raggedright\arraybackslash}p{0.20\textwidth}r>{\raggedright\arraybackslash}p{0.20\textwidth}@{}}
    \toprule
    Benchmark & Source snapshot & Eligible pool & Main $n$ & Length strata \\
    \midrule
    BABILong & QA1--QA5, revision \path{ee0d5887} & 300 independent & 300 & 1K / 2K / 4K; 20 per QA--length pair \\
    HotpotQA & LongBench v1, revision \path{5e628be4} & 199 rows; 199 documents & 100 & Four token-length quartiles \\
    2WikiMQA & LongBench v1, revision \path{5e628be4} & 192 rows; 192 documents & 100 & Four token-length quartiles \\
    MultiFieldQA-en & LongBench v1, revision \path{5e628be4} & 139 rows; 105 documents & 85 & Four token-length quartiles; 15-sample shortfall \\
    \bottomrule
  \end{tabular}
\end{table}

BABILong selection balances QA1--QA5 and the three source lengths. The LongBench tasks use token-length quartiles. All 585 evaluated samples are distinct at the story or source-document level. BABILong uses accuracy, and the other three tasks use token F1. Each request uses greedy decoding with a 48-token output limit. Failed requests receive zero task score.

Table~\ref{tab:app-prefix-lengths} reports actual prefix lengths after official chat encoding. Median question-suffix lengths are 11, 24, 22, and 19 tokens for BABILong, HotpotQA, 2WikiMQA, and MultiFieldQA-en, respectively. The benchmark-level TTFT results therefore compare workloads with different input-length distributions.

\begin{table}[!htbp]
  \centering
  \caption{Tokenized prefix lengths in the 585-sample main study. Q1 and Q3 are the first and third quartiles. Question suffixes are excluded.}
  \label{tab:app-prefix-lengths}
  \small
  \begin{tabular}{@{}lrrrrrr@{}}
    \toprule
    Benchmark & $n$ & Min & Q1 & Median & Q3 & Max \\
    \midrule
    BABILong & 300 & 571 & 735 & 1,712 & 3,539 & 3,852 \\
    HotpotQA & 100 & 2,142 & 11,054 & 14,320 & 15,892 & 16,507 \\
    2WikiMQA & 100 & 2,580 & 5,041 & 6,369 & 8,530 & 15,783 \\
    MultiFieldQA-en & 85 & 2,703 & 4,827 & 7,666 & 9,030 & 14,636 \\
    \bottomrule
  \end{tabular}
\end{table}

TTFT is measured with the source cache already prepared, from the target continuation request to the first non-empty generated token. It includes cache lookup, loading, restoration, and the target's remaining computation. Source prefill, cache promotion, target startup, and model loading are excluded. Target uses a cold request without prefix-cache hits. Appendix~\ref{app:main-results} gives the complete quality and latency results.

\FloatBarrier
\subsection{Behavioral and Mechanistic Analysis}
\label{app:diagnostic-data}

The Flash-update state traces, token-score comparisons, and interventions use independent 4K contexts. Table~\ref{tab:app-diagnostic-cohorts} summarizes the primary diagnostic cohorts. The main and confirmation contexts are distinct. All runs use greedy decoding, with an eight-token output limit for retrieval and a 256-token limit for QA to support content and stopping analysis.

\begin{table}[!htbp]
  \centering
  \caption{Primary diagnostic cohorts. Retrieval counts are queries and QA counts are examples.}
  \label{tab:app-diagnostic-cohorts}
  \small
  \setlength{\tabcolsep}{4pt}
  \begin{tabularx}{\linewidth}{@{}>{\raggedright\arraybackslash}Xr>{\raggedright\arraybackslash}p{0.30\linewidth}@{}}
    \toprule
    Cohort & $n$ & Independent unit \\
    \midrule
    Main retrieval diagnostics & 96 & 32 contexts \\
    Retrieval confirmation & 48 & 16 contexts \\
    Natural-QA diagnostics & 89 & 89 examples \\
    \bottomrule
  \end{tabularx}
\end{table}

The 89 QA examples comprise 39 HotpotQA and 50 2WikiMQA examples. Each task contributes 20 random examples selected before the new runs from its original 100-example pool. Other cases examine earlier loss, gain, and stability patterns. Frequency summaries use only the random subsets.

Exploratory retrieval analyses use 33 unique queries: 24 position-selected queries and 12 historical cases, with three overlaps. The four selected retrieval losses in Appendix~\ref{app:crossed} come from the main and exploratory cohorts, with overlapping cases counted once.

Figure~\ref{fig:drift-and-margins} uses additional 4K NIAH comparisons. The Flash-0731 traces contain 12 behavior-selected queries from seven histories: four improved, four degraded, and four correct in both runs. The Flash-Base traces contain four improved queries from three histories. NMSE aggregates all query positions and features at each measurement site. These selected traces illustrate internal trajectories rather than estimate population frequencies. The margin panels instead include the same 192 depth queries from eight histories for each receiver, with identical prefixes, question suffixes, and gold answers across receivers. All 192 queries enter the outcome counts and margin summaries, including the single extreme query outside each displayed margin range. Appendix~\ref{app:runtime} specifies their prefill settings.

Figure~\ref{fig:indexer-long-context} uses four Flash-to-Base retrieval-recovery queries at each of 4K, 8K, 16K, and 32K, from three, four, four, and three histories, respectively. The 4K queries are the same as the Flash-Base traces in Figure~\ref{fig:drift-and-margins}. Each length uses its own selected queries, all incorrect under Target and correct under KV Reuse by answer-prefix matching. At the final query position, we compare the actual selected sets at the 21 Indexer layers, numbered 2--42 in steps of two. Overlap is $r=|S_T\cap S_R|/|S_T|$, where $S_T$ and $S_R$ are the Target and KV Reuse selections. With a budget of $k=512$ and $N$ valid candidates, the uniform-random reference is $b=\min(k,N)/N$, and adjusted overlap is $(r-b)/(1-b)$. Heatmap cells average overlap over the four queries. Layer-band curves average adjusted overlap over queries and the indicated layers. These summaries describe the selected recovery cases.

\FloatBarrier
\subsection{Functional Cache Divergence}
\label{app:fcd-protocol}

FCD compares full-vocabulary, temperature-one next-token distributions in FP32 using the definition in Equation~\ref{eq:fcd-pointwise}. The NIAH evaluation uses the same 1,152 queries as the six-length Flash-update sweep. Each comparison is made immediately after the common question suffix, before supplying any answer token. It yields one FCD value per query, regardless of the eventual response length.

FineWeb-Edu~\citep{penedo2024fineweb} uses \texttt{sample-10BT} at revision \path{87f09149}. We select the first 64 unique documents with at least 100,000 characters and enough tokens for a 32K prefix plus the continuation. The same documents are used at all six lengths. After caching $L$ tokens, both target runs consume the same original-document tokens $t_1,\ldots,t_{32}$ and predict $t_2,\ldots,t_{33}$, yielding 32 paired positions per document and 12,288 over all six lengths. Identical supplied tokens keep subsequent inputs aligned when predictions differ.

NIAH means weight histories equally after averaging their query values. FineWeb-Edu means weight documents equally after averaging their prediction positions. Confidence intervals use 10,000 history- or document-cluster bootstrap resamples. Supplementary statistics are reported in Appendix~\ref{app:fcd-outcomes}.

\FloatBarrier

\section{Cache Handoff and Runtime Controls}
\label{app:implementation}

\subsection{Composite Cache and Page Validation}
\label{app:cache}

Both handoffs transfer the architecture-specific state needed for continuation: the KV anchor, sliding-window attention state (SWA), compression state C4 and its indexer state, compression state C128, and the two associated state bundles. The detailed retrieval, generation, and intervention analyses transfer this complete prefix state at the boundary between history $P$ and question $Q$. The serving benchmark instead transfers complete pages from file-backed storage, after which the target processes any unaligned prefix tail before $Q$.

The SGLang~\citep{zheng2024sglang} experiments use 256-token storage pages and 4,096-token chunked prefill. For a prefix of $P$ tokens, the expected transferable extent is
\[
  P_{\mathrm{storage}} = 256\left\lfloor P/256 \right\rfloor .
\]
Before continuation, the harness verifies file presence, size, and SHA-256 for every required page component. The native request must have zero cache hits; a direct request must hit the expected storage extent with zero device/host hits. The same checks are repeated after the run. A partial or non-storage hit is a structural failure rather than a successful direct result.

\subsection{Runtime Matrix}
\label{app:runtime}

The length sweeps and detailed diagnostics use different prefill paths. Flash prefill is chunked at 512 tokens for the short sweep and 1,024 for 8K--32K, while Qwen uses 512-token chunks. The GLM sweep combines 512--4K runs with 8K--32K runs using 1,024-token prefill chunks and a different cache-storage layout. The serving benchmark transfers complete storage pages, whereas the formal 96-query diagnostics use full-prefix prefill. For Figure~\ref{fig:drift-and-margins}, the Flash-0731 comparison uses 512-token prefill chunks and the Flash-Base comparison uses 1,024-token chunks. Each paired run uses the same chunk size, and the complete question suffix is processed in one forward pass after prefix-state transfer. Table~\ref{tab:app-runtime} summarizes the runtime and handoff settings. Generation settings are given in Appendix~\ref{app:data}.

The Flash-to-Base Indexer comparisons in Figure~\ref{fig:indexer-long-context} use 1,024-token prefill chunks throughout. The 4K run uses PyTorch 2.10.0 and the 8K--32K runs use PyTorch 2.11.0, all with Transformers 5.20.1 and CUDA 12.9. Target and KV Reuse share execution settings within each length.

Paired comparisons share execution settings within each experiment. We compute 95\% confidence intervals by resampling independent histories for retrieval and source examples for QA, using 10,000 bootstrap replicates for the length sweeps and 2,000 for the serving and detailed analyses. FCD uses 512-token prefill chunks through 4K and 1,024-token chunks at longer lengths for NIAH, and 1,024-token chunks throughout for natural text. Its aggregation and confidence intervals are described in Appendix~\ref{app:fcd-protocol}.

\begin{table}[!htbp]
  \centering
  \caption{Runtime and cache-handoff settings by experiment family.}
  \label{tab:app-runtime}
  \footnotesize
  \begin{tabular}{@{}>{\raggedright\arraybackslash}p{0.22\textwidth}>{\raggedright\arraybackslash}p{0.32\textwidth}>{\raggedright\arraybackslash}p{0.36\textwidth}@{}}
    \toprule
    Experiment family & Execution & Cache and prefill settings \\
    \midrule
    585-sample main study & SGLang commit \path{1fdd6c89}; eight H100 GPUs; TP=8; file HiCache & 4,096-token chunks; 256-token pages; \texttt{write\_through} \\
    NIAH readout & Transformers and PyTorch on eight H100 GPUs & Complete prefix-state transfer \\
    Flash ridge-mapper comparison & Transformers 5.20.1; PyTorch 2.11.0+cu129; eight H100 GPUs & Shared FP8 Triton path for all four configurations; 512-token chunks through 4K, 1,024 at 8K--32K \\
    GLM additional-update NIAH & SGLang; TP=2, PP=8 across two GPU nodes & File-backed cache handoff \\
    Pro additional-update NIAH & PyTorch reference runtime; separate source and target workers & Full prefill; complete prefix-cache handoff \\
    V4.1 additional-update NIAH & Official reference runtime; TP=8 & Complete prefix-state restoration \\
    Activation and stopping analyses & Transformers and PyTorch on eight H100 GPUs & Full-prefix prefill for formal cohorts; chunked prefill for Figure~\ref{fig:drift-and-margins} \\
    \bottomrule
  \end{tabular}
\end{table}

For the V4.1-Flash extension, 64K--512K prefixes use 8,192-token continuation-chunk prefill with fixed projection blocks, complete historical compressed states, and absolute positions. The earlier 512--32K results use the prior full-prefill path. Native and reuse runs share execution settings within each length. At 256K and 512K, the common query-suffix prefix is cached separately for each history and configuration; its first query is checked against the complete suffix path for identical logits and output tokens. Cross-length trends therefore describe the recorded settings, including the 512K construction difference, rather than isolate length as the only changing factor.

\section{Extended Results}
\label{app:results}

\subsection{Complete Benchmark Results}
\label{app:main-results}

\begin{table}[!htbp]
  \centering
  \caption{Task quality on the 585-sample benchmark study. Scores are percentages. $\Delta$ is direct minus target-native, in percentage points; intervals use 2,000 document/story-cluster bootstrap resamples. Score $\downarrow/\uparrow$ counts paired score decreases and increases. Flips report native-success$\rightarrow$direct-failure and native-failure$\rightarrow$direct-success. Failures are native/direct and score zero.}
  \label{tab:app-main-quality}
  \footnotesize
  \setlength{\tabcolsep}{3pt}
  \begin{tabular}{@{}lrlrrrp{0.21\textwidth}ccc@{}}
    \toprule
    Benchmark & $n$ & Metric & Target & Direct & Source & $\Delta$ [95\% CI] & Score $\downarrow/\uparrow$ & Flips & Fail. \\
    \midrule
    BABILong & 300 & Acc. & 59.0 & 58.7 & 61.0 & $-0.3\,[-4.3,+3.7]$ & 20/19 & 20/19 & 0/0 \\
    HotpotQA & 100 & Token F1 & 69.3 & 66.3 & 69.4 & $-3.0\,[-6.6,-0.1]$ & 8/6 & 5/0 & 0/0 \\
    2WikiMQA & 100 & Token F1 & 61.7 & 54.7 & 58.9 & $-7.0\,[-12.7,-1.5]$ & 19/3 & 13/3 & 0/2 \\
    MultiFieldQA-en & 85 & Token F1 & 45.9 & 46.2 & 49.8 & $+0.3\,[-3.1,+3.7]$ & 20/19 & 1/2 & 5/5 \\
    \bottomrule
  \end{tabular}
\end{table}

The source-native score is descriptive context for the representation producer. It is not a target-quality baseline, and no aggregate score is formed across benchmarks because their metrics and score distributions differ.

\begin{table}[!htbp]
  \centering
  \caption{Prepared-cache TTFT. Only valid target/direct pairs enter the sample count and percentiles. The speedup is computed per sample as target-native/direct TTFT and summarized by its median; it is not the ratio of the two marginal medians.}
  \label{tab:app-main-ttft}
  \footnotesize
  \setlength{\tabcolsep}{4pt}
  \begin{tabular}{@{}lrrrrr@{}}
    \toprule
    Benchmark & Pairs & Native p50/p95 & Direct p50/p95 & Speedup p50 & Direct slower \\
    \midrule
    BABILong & 300/300 & 153.2/286.5 & 192.2/254.4 & 0.855$\times$ & 201 \\
    HotpotQA & 100/100 & 1,792.3/8,101.1 & 355.9/392.2 & 5.083$\times$ & 1 \\
    2WikiMQA & 98/100 & 543.5/1,712.7 & 267.3/524.2 & 2.097$\times$ & 3 \\
    MultiFieldQA-en & 80/85 & 759.1/1,480.8 & 248.2/762.1 & 2.512$\times$ & 6 \\
    \bottomrule
  \end{tabular}
\end{table}

Latency follows the protocol in Appendix~\ref{app:datasets}. Each request was timed once, so the measurements describe these runs rather than service-level latency.

\subsection{Length-Stratified Results}
\label{app:length-results}

\begin{table}[!htbp]
  \centering
  \caption{Quality and prepared-cache TTFT by actual input-length group. Scores and confidence intervals are percentages. BABILong groups are source-defined lengths; other groups are quartiles of tokenized prefixes.}
  \label{tab:app-length-results}
  \scriptsize
  \setlength{\tabcolsep}{2.5pt}
  \begin{tabular}{@{}lrrrrp{0.20\textwidth}rcr@{}}
    \toprule
    Group & $n$ & Fail. & Target/Direct & $\Delta$ [95\% CI] & Score $\downarrow/\uparrow$ & Pairs & Speedup \\
    \midrule
    2Wiki q1 & 25 & 0/0 & 74.6/66.9 & $-7.8\,[-24.6,+9.1]$ & 7/2 & 25/25 & 1.80$\times$ \\
    2Wiki q2 & 25 & 0/1 & 59.5/56.8 & $-2.7\,[-10.5,+4.3]$ & 3/1 & 24/25 & 2.09$\times$ \\
    2Wiki q3 & 25 & 0/0 & 56.2/47.5 & $-8.8\,[-18.0,-1.3]$ & 4/0 & 25/25 & 2.21$\times$ \\
    2Wiki q4 & 25 & 0/1 & 56.4/47.5 & $-8.9\,[-19.2,-0.6]$ & 5/0 & 24/25 & 4.52$\times$ \\
    \addlinespace
    BABILong 1K & 100 & 0/0 & 67.0/67.0 & $0.0\,[-7.0,+8.0]$ & 7/7 & 100/100 & 0.86$\times$ \\
    BABILong 2K & 100 & 0/0 & 57.0/60.0 & $+3.0\,[-3.0,+9.0]$ & 4/7 & 100/100 & 0.79$\times$ \\
    BABILong 4K & 100 & 0/0 & 53.0/49.0 & $-4.0\,[-11.0,+3.0]$ & 9/5 & 100/100 & 1.23$\times$ \\
    \addlinespace
    Hotpot q1 & 25 & 0/0 & 70.1/68.5 & $-1.6\,[-10.6,+6.0]$ & 2/1 & 25/25 & 2.40$\times$ \\
    Hotpot q2 & 25 & 0/0 & 68.9/63.3 & $-5.6\,[-14.1,+0.0]$ & 3/1 & 25/25 & 5.55$\times$ \\
    Hotpot q3 & 25 & 0/0 & 69.6/65.0 & $-4.6\,[-12.9,+0.5]$ & 2/2 & 25/25 & 4.88$\times$ \\
    Hotpot q4 & 25 & 0/0 & 68.5/68.4 & $-0.1\,[-1.2,+0.7]$ & 1/2 & 25/25 & 10.51$\times$ \\
    \addlinespace
    MultiField q1 & 22 & 4/4 & 43.5/43.8 & $+0.3\,[-0.2,+1.0]$ & 1/3 & 18/22 & 1.23$\times$ \\
    MultiField q2 & 21 & 1/1 & 44.5/48.8 & $+4.3\,[-5.9,+14.4]$ & 4/9 & 20/21 & 2.26$\times$ \\
    MultiField q3 & 21 & 0/0 & 44.3/46.9 & $+2.6\,[-3.1,+10.2]$ & 6/3 & 21/21 & 3.54$\times$ \\
    MultiField q4 & 21 & 0/0 & 51.5/45.6 & $-5.9\,[-12.3,-0.5]$ & 9/4 & 21/21 & 4.11$\times$ \\
    \bottomrule
  \end{tabular}
\end{table}

These strata are descriptive. They were not designed to estimate a single causal slope of speedup against prefix length, and input data and serving overhead differ across benchmarks.

\subsection{Same-Version Controls and Sensitivity Checks}
\label{app:controls}

\begin{table}[!htbp]
  \centering
  \caption{Target-self controls in the 585-sample study. ``First'' and ``full'' report first-token and complete-recorded-sequence agreement, respectively. Scores show mean self minus native and direct minus self gaps.}
  \label{tab:app-self-controls}
  \footnotesize
  \begin{tabular}{@{}lrrrrr@{}}
    \toprule
    Benchmark & $n$ & First match & Full match & Self$-$Native & Direct$-$Self \\
    \midrule
    BABILong & 300 & 265 & 255 & $-0.013$ & $+0.010$ \\
    HotpotQA & 100 & 95 & 87 & $-0.010$ & $-0.020$ \\
    2WikiMQA & 100 & 94 & 84 & $-0.013$ & $-0.058$ \\
    MultiFieldQA-en & 85 & 82 & 52 & $-0.005$ & $+0.008$ \\
    \bottomrule
  \end{tabular}
\end{table}

The dedicated 129-query retrieval and 89-example natural-QA analysis paths match across native, restored-self, and fresh-process self. The 585-sample serving study shows the same-version differences in Table~\ref{tab:app-self-controls}, including complete-sequence agreement of 52/85 on MultiFieldQA-en. Its direct--native comparisons therefore combine cross-version effects with variation observed under the serving handoff.

Seven observations in the original benchmark table failed a storage-tier validity condition. A post hoc retest started the target separately for each affected sample and restored copied, verified pages. It changed two rows as follows; the original table remains the primary result.

\begin{table}[!htbp]
  \centering
  \caption{Post hoc cache-tier sensitivity retest. Scores and score changes use each task's primary metric. The retest isolates seven rows with device/host cache interference and is not pooled with the original main result.}
  \label{tab:app-cache-tier-retest}
  \footnotesize
  \setlength{\tabcolsep}{4pt}
  \begin{tabular}{@{}llrrr@{}}
    \toprule
    Benchmark & Valid pairs (original$\rightarrow$retest) & Target/direct & $\Delta$ [95\% CI] & Speedup \\
    \midrule
    2WikiMQA & 98/100 $\rightarrow$ 100/100 & 0.617/0.548 & $-0.069\,[-0.126,-0.013]$ & 2.097$\times$ \\
    MultiFieldQA-en & 80/85 $\rightarrow$ 85/85 & 0.501/0.506 & $+0.005\,[-0.036,+0.047]$ & 2.390$\times$ \\
    \bottomrule
  \end{tabular}
\end{table}

\subsection{Output Divergence}
\label{app:output-review}

The preselected random QA subsets provide additional paired quality measurements for the detailed diagnostics discussed in Section~\ref{sec:sequence-agreement}. Table~\ref{tab:qa-random-quality} reports the full-response F1 scores for 20 examples per task.

\begin{table}[!htbp]
  \centering
  \caption{Full-response F1 on the preselected random QA subsets. Loss, unchanged, and gain count each example's paired score change. Intervals use 2,000 bootstrap resamples of source examples.}
  \label{tab:qa-random-quality}
  \small
  \setlength{\tabcolsep}{5pt}
  \begin{tabular}{@{}lcccl@{}}
    \toprule
    Task ($n=20$ each) & Target F1 & KV Reuse F1 & Loss / unchanged / gain & Change [95\% CI] \\
    \midrule
    HotpotQA & 0.6553 & 0.6683 & 1 / 18 / 1 & $+0.0129\;[-0.0088,+0.0476]$ \\
    2WikiMQA & 0.6382 & 0.5571 & 5 / 13 / 2 & $-0.0811\;[-0.2131,+0.0435]$ \\
    \bottomrule
  \end{tabular}
\end{table}

\begin{table}[!htbp]
  \centering
  \caption{First-divergence analysis for the four cross-benchmark QA tasks. Only samples with successful native/direct pairs enter this table. ``Later'' means the first token matched and the recorded sequences still differed.}
  \label{tab:app-divergence}
  \footnotesize
  \begin{tabular}{@{}lrrrrrr@{}}
    \toprule
    Benchmark & Pairs & Identical & First token & Later & Later score $\downarrow/=/\!\!\uparrow$ & Later EM $1\!\rightarrow\!0$ \\
    \midrule
    BABILong & 300 & 199 & 87 & 14 & 0/12/2 & 0 \\
    HotpotQA & 100 & 70 & 6 & 24 & 7/11/6 & 5 \\
    2WikiMQA & 98 & 59 & 11 & 28 & 12/15/1 & 9 \\
    MultiFieldQA-en & 80 & 33 & 10 & 37 & 16/6/15 & 0 \\
    \bottomrule
  \end{tabular}
\end{table}

\subsection{Functional Divergence by Retrieval Outcome}
\label{app:fcd-outcomes}

This subsection supplements the FCD comparison in Section~\ref{sec:fcd-tasks}. Data selection, matched prediction positions, and aggregation are described in Appendix~\ref{app:fcd-protocol}.

On FineWeb-Edu, mean reuse-minus-target negative log likelihood on the supplied next tokens is positive at every length, ranging from 0.036 to 0.082 nats per token. At 32K, mean FCD is 3.51 (95\% CI: $[3.11,3.94]$), with 86.62\% top-1 agreement.

The NIAH FCD distribution is skewed at short prefixes. At 512 tokens, mean FCD is 2.87 (95\% CI: $[1.12,5.11]$), while median FCD is 0.035. At 16K, the mean reaches 24.11 ($[20.05,28.18]$). At 32K, the mean is 30.62 ($[25.71,35.51]$) and the median is 23.51. Target/Reuse top-1 agreement falls from 96.35\% at 512 to 68.75\% at 32K.

FCD measures deviation from target-native predictions without assigning a beneficial or harmful outcome. Table~\ref{tab:fcd-outcomes} groups the 16K and 32K retrieval measurements by the previously recorded answer-prefix outcomes on the same query IDs. At 16K, the mean FCD of the 36 improved answers is 66.42, exceeding the 44.34 mean for the eight degraded answers. At 32K, gains and losses both have high means, 63.23 and 66.15, compared with 9.99 among the 36 queries answered correctly by both configurations. Large functional changes can therefore accompany correction of a target error as well as loss of a correct answer. Both-incorrect outcomes also occur without the largest divergences: the 119 such queries at 32K have mean FCD 26.40, below the gain and loss groups.

\begin{table}[htbp]
  \centering
  \caption{FCD by paired retrieval outcome at 16K and 32K. Categories use the frozen answer-prefix evaluation; larger FCD denotes greater distributional change. Counts and conditional summaries describe the observed groups.}
  \label{tab:fcd-outcomes}
  \small
  \setlength{\tabcolsep}{7pt}
  \begin{tabular}{@{}llrrr@{}}
    \toprule
    Prefix & Target $\rightarrow$ Reuse outcome & Queries & Mean FCD & Median FCD \\
    \midrule
    16K & Correct $\rightarrow$ correct & 89 & 4.56 & 1.04 \\
        & Incorrect $\rightarrow$ correct & 36 & 66.42 & 73.12 \\
        & Correct $\rightarrow$ incorrect & 8 & 44.34 & 33.09 \\
        & Incorrect $\rightarrow$ incorrect & 59 & 25.03 & 21.79 \\
    \addlinespace
    32K & Correct $\rightarrow$ correct & 36 & 9.99 & 6.99 \\
        & Incorrect $\rightarrow$ correct & 24 & 63.23 & 65.31 \\
        & Correct $\rightarrow$ incorrect & 13 & 66.15 & 71.50 \\
        & Incorrect $\rightarrow$ incorrect & 119 & 26.40 & 21.95 \\
    \bottomrule
  \end{tabular}
\end{table}

The FCD run is a separate measurement campaign, not a replay of the full generation evaluation. Its first-token predictions differ from the frozen behavior run in 60 of 2,304 arm-level comparisons, all at 512--4K. The 16K/32K groups in Table~\ref{tab:fcd-outcomes} have no such first-token mismatches. Prefill and cache-restoration settings are given in Appendix~\ref{app:runtime}. This FCD campaign has no target-self restoration baseline, so a separate estimate of distributional variation introduced by restoration is unavailable. The exact self-restoration results of the full-prefix diagnostic experiments in Section~\ref{sec:internal-diagnostics} apply to those experiments, not automatically to this campaign.

These results show that functional divergence varies across tasks and can accompany both improved and degraded answers in the measured samples. They do not validate FCD as a predictor of harmful reuse on unseen inputs. Task outcomes and generation fidelity remain necessary alongside FCD; a single average divergence does not summarize which examples change or whether those changes are useful.

\section{NIAH Readout and Mechanism Details}
\label{app:mechanism}

\subsection{Flash-to-Base Readout Diagnostics}
\label{app:base-readout}

Answer-prefix accuracy permits leading whitespace and short formatting tokens. First-token correctness, strict exact match, gold-token rank, and top-10 membership separate content readout from response formatting.

\begin{table}[!htbp]
  \centering
  \caption{Flash-to-Base NIAH readout diagnostics. Gold rank is over the full first-token vocabulary; lower rank is better. The prefix, first-token, and exact columns are counts out of 192.}
  \label{tab:app-base-readout}
  \scriptsize
  \setlength{\tabcolsep}{3pt}
  \begin{tabular}{@{}rrlrrrrrr@{}}
    \toprule
    Prefix & Arm & Prefix & First token & Exact & Median rank & Mean prob. & Top-10 (\%) \\
    \midrule
    512 & Native & 0 & 0 & 0 & 2151.5 & 0.0019 & 7.8 \\
    512 & Direct & 169 & 0 & 169 & 2.0 & 0.1056 & 97.9 \\
    512 & Source & 190 & 190 & 186 & 1.0 & 0.9841 & 100.0 \\
    2,048 & Native & 0 & 0 & 0 & 18508.0 & $2.5\!\times\!10^{-5}$ & 0.0 \\
    2,048 & Direct & 175 & 15 & 174 & 2.0 & 0.1792 & 99.0 \\
    2,048 & Source & 187 & 187 & 186 & 1.0 & 0.9621 & 99.5 \\
    4,096 & Native & 0 & 0 & 0 & 16466.0 & $1.4\!\times\!10^{-5}$ & 0.0 \\
    4,096 & Direct & 157 & 14 & 155 & 2.0 & 0.1847 & 97.9 \\
    4,096 & Source & 183 & 183 & 183 & 1.0 & 0.9402 & 99.5 \\
    \bottomrule
  \end{tabular}
\end{table}

The Direct arm generally does not select the value as its literal first token, particularly at 2K and 4K, yet strict match after stripping whitespace is almost identical to answer-prefix accuracy. This pattern is consistent with a readable value following a formatting token. To test whether the eight-token limit caused the native failure, a continuation probe appended the native arm's own eight generated template tokens and inspected the following gold-token distribution. At 512 tokens, gold is top-1 in 1/192 cases, top-10 in 35.9\%, and has median rank 26; at 4,096, it is top-1 in 2/192, top-10 in 25.5\%, and has median rank 30. The corresponding Direct median rank is 2. The result therefore does not reduce solely to the output limit.

\subsection{Crossed Input--Cache Intervention}
\label{app:crossed}

For the 96-query retrieval cohort, target attention at layers 27, 37, and 42 is evaluated in four copies: target input with target cache, target input with source cache, source-path input with target cache, and source-path input with source cache. Layer weights, token positions, and index-selection recomputation are held fixed as specified in the main text. Table~\ref{tab:app-crossed} reports median normalized differences at the prediction position.

\begin{table}[!htbp]
  \centering
  \caption{Crossed input--cache intervention at the prediction position. All values are normalized attention-output differences from the target-native result, median over 96 queries. Columns are not additive because changing both factors also changes their interaction.}
  \label{tab:app-crossed}
  \small
  \begin{tabular}{@{}lrrr@{}}
    \toprule
    Layer (zero-based) & \shortstack{Replace local\\KV cache} & \shortstack{Replace incoming\\hidden state} & Replace both \\
    \midrule
    27 & 0.012 & 0.142 & 0.139 \\
    37 & 0.019 & 0.473 & 0.530 \\
    42 & 0.005 & 0.064 & 0.077 \\
    \bottomrule
  \end{tabular}
\end{table}

For the four selected retrieval losses, the replacement outcomes are summarized in Table~\ref{tab:app-cache-versus-output}. The cases combine the exploratory and main cohorts, with overlapping roles counted once.

\begin{table}[!htbp]
  \centering
  \caption{Recovery of four selected retrieval losses after replacing native attention outputs or cache components. Layer numbers are zero-based.}
  \label{tab:app-cache-versus-output}
  \small
  \begin{tabular}{@{}lcc@{}}
    \toprule
    Replacement in the direct path & Layers & Answers restored \\
    \midrule
    Native attention outputs & 36--42 & 4/4 \\
    Complete native prefix caches & 36--42 & 1/4 \\
    Complete native prefix caches & 0--27 & 2/4 \\
    Native compressor state bundles & 0--27 & 3/4 \\
    \bottomrule
  \end{tabular}
\end{table}

The complete cache replacement over all layers reproduces the native sequence on all 96 queries. Replacing all suffix attention outputs reproduces all 96 first tokens but only 90 complete sequences, which is why output patches are interpreted as local causal tests rather than deployable cache repair.

\subsection{Matched Perturbations and Stopping}
\label{app:perturbation}

The confirmation cohort contains 48 queries from 16 contexts not used by the main retrieval cohort. At layer 37, a natural attention-output difference is injected at full strength and compared with independent per-token random directions and a shared structured rotation. Four repetitions are averaged per query; BF16-rounded perturbation norms are matched at each position within 0.2\%.

The mean absolute changes in the native winner's dynamic margin are 0.693 for the natural patch, 0.624 for independent random directions, and 0.611 for the shared rotation. The random-minus-natural paired effects are $-0.068$ with 95\% interval $[-0.204,+0.055]$ for independent directions and $-0.081$ with interval $[-0.249,+0.078]$ for the rotation. Both intervals span zero. At layer 42, the corresponding mean margin changes are 1.724, 0.545, and 0.533, while the natural patch changes the selected first token on one of 48 queries. These patches act during suffix processing with the other layers' caches native, so they measure local sensitivity rather than the full direct-reuse trajectory.

For this supplementary stopping probe, we use $m_{\mathrm{stop}}=z_{\mathrm{EOS}}-\max_{v\ne\mathrm{EOS}}z_v$, the EOS token's lead over the strongest continuation token. A positive margin favors stopping. Target and KV Reuse continue from the same first reference-answer tokens without consuming EOS. On the 20-example random HotpotQA subset, the mean stopping-margin change is $-1.131$ with 95\% interval $[-1.581,-0.700]$ and decreases on 17/20 examples. On the 20-example random 2WikiMQA subset, it is $-1.150$ with interval $[-1.850,-0.488]$ and decreases on 15/20 examples. The intervals describe these preselected subsets from the original 100-example task pools.

\end{document}